\documentclass[floatfix,nofootinbib,a4paper,aps,prd,twocolumn,preprintnumbers]{revtex4-1}  
\usepackage{epsfig} 
\usepackage{axodraw} 
\usepackage{amsmath} 
\usepackage{amssymb} 
\usepackage{amsfonts} 
\usepackage{graphicx} 
\usepackage{graphics,subfigure} 
\usepackage{axodraw} 
\usepackage{float} 
\usepackage{booktabs} 
\usepackage{color} 
\usepackage{geometry} 
\usepackage{rotating} 
\usepackage{units} 
\usepackage{slashed} 
\usepackage{appendix}
 
\newcommand{\lsim}{\raisebox{-0.13cm}{~\shortstack{$<$ \\[-0.07cm] $\sim$}}~}

\newcommand{\bea}{\begin{eqnarray}} 
\newcommand{\eea}{\end{eqnarray}} 
\newcommand{\beq}{\begin{equation}} 
\newcommand{\eeq}{\end{equation}} 
\newcommand{\beqa}{\begin{eqnarray}} 
\newcommand{\eeqa}{\end{eqnarray}} 
\newcommand{\bit}{\begin{itemize}} 
\newcommand{\eit}{\end{itemize}}

\newcommand{\met}{\ensuremath{\slashed{E}_T}}

\newbox\charbox 
\newbox\slabox 
\def\s#1{{      
    \setbox\charbox=\hbox{$#1$} 
    \setbox\slabox=\hbox{$/$} 
    \dimen\charbox=\ht\slabox 
    \advance\dimen\charbox by -\dp\slabox 
    \advance\dimen\charbox by -\ht\charbox 
    \advance\dimen\charbox by \dp\charbox 
    \divide\dimen\charbox by 2 
    \raise-\dimen\charbox\hbox to \wd\charbox{\hss/\hss} 
    \llap{$#1$} 
}}

\begin{document} 
  
\title{Minimal Natural Supersymmetry after the LHC8} 
 
\author{Manuel Drees} 
\email[]{drees@th.physik.uni-bonn.de} 
\affiliation{BCTP and Physics Institute, University of Bonn, Bonn, Germany } 
  
\author{Jong Soo Kim} 
\email[]{jong.kim@csic.es} 
\affiliation{Instituto de F\'{\i}sica Te\'{o}rica UAM/CSIC, Madrid, Spain}

\begin{abstract}
 
  In this work, we present limits on natural supersymmetry scenarios
  based on searches in data taken during run 1 of the LHC. We consider
  a set of 22000 model points in a six dimensional parameter
  space. These scenarios are minimal in the sense of only keeping
  those superparticles relatively light that are required to cancel
  the leading quadratically divergent quantum corrections (from the
  top and QCD sector) to the Higgs mass in the Standard Model. The
  resulting mass spectra feature higgsinos as the lightest
  supersymmetric particle, as well as relatively light third
  generation $SU(2)$ doublet squarks and $SU(2)$ singlet stops and
  gluinos while assuming a Standard Model like Higgs boson. All
  remaining supersymmetric particles and Higgs bosons are assumed to
  be decoupled. We check each parameter set against a large number of
  LHC searches as implemented in the public code {\tt
    CheckMATE}. These searches show a considerable degree of
  complementarity, i.e. in general many searches have to be considered
  in order to check whether a given scenario is allowed. We delineate
  allowed and excluded regions in parameter space. For example, we
  find that all scenarios where either $m_{\tilde t_1} < 230$ GeV or
  $m_{\tilde g} < 440$ GeV are clearly excluded, while all model
  points where $m_{\tilde t_1} > 660$ GeV and $m_{\tilde g} > 1180$
  GeV remain allowed.
\end{abstract} 
 
\preprint{IFT-UAM/CSIC-15-104}

\maketitle 
  
\section{Introduction} 
\label{sec:intro} 

Supersymmetry (SUSY) is one of the best motivated extensions of the
standard model (SM); it stabilizes the gauge hierarchy against
radiative corrections, and allows one--step unification of the gauge
couplings of the SM \cite{susy}. Assuming a discrete symmetry like
$R-$parity \cite{Ibanez:1991pr}, the lightest supersymmetric particle
(LSP) is stable, and can make a good Dark Matter candidate
\cite{susy}. This symmetry also implies that supersymmetric particles
can only be produced in pairs and subsequently decay into SM particles
and the LSP. Since the LSP escapes detection at both multipurpose
detectors ATLAS and CMS, supersymmetric particle production can give
rise to large missing transverse momentum, accompanied by high
momentum jets and leptons.

The LHC detectors started to take data in 2010 at the center of mass
energy of 7 TeV and both collected about 5 fb$^{-1}$ of integrated
luminosity. In 2012, the center of mass energy was increased to
$\sqrt{s}=8$ TeV and at the end of the run, both ATLAS and CMS
recorded about 20 fb$^{-1}$ of data. Unfortunately no significant
excess above the SM expectation has been found, although a few $2$ to
$3\sigma$ anomalies have been found, which can be explained in the
framework of supersymmetry \cite{Kim:2014eva,Allanach:2014lca}. The
null results of both experiments have been translated into strict
bounds on extensions of the SM. These extensions include
supersymmetric models with simplifying assumptions on the soft
breaking sector such as in the mSUGRA/CMSSM model
\cite{Chamseddine:1982jx}, as well as simplified models containing
only a few supersymmetric particles and couplings
\cite{Alwall:2008va}. For example, assuming degenerate squark and
gluino masses as well as a light neutralino LSP, squarks and gluinos
below 1.8 TeV are now excluded \cite{Aad:2015iea}.

These search limits, together with the discovery of a relatively heavy
SM--like Higgs boson \cite{Aad:2015zhl}, put some strain on a
supersymmetric solution of the finetuning problem. As well known, in
theories with exact supersymmetry the Higgs mass is completely
unaffected by loop corrections. Once soft breaking terms are
introduced, quadratically divergent corrections continue to cancel,
unlike in the SM, but there are corrections to the squared Higgs mass
parameters in the Lagrangian that scale with (combinations of) squares
of these soft masses. At the same time corrections to the {\em
  physical} mass of the lightest CP--even neutral Higgs boson, which
at the tree--level is lighter than the $Z$ boson, only scale
logarithmically with the soft SUSY breaking parameters, in particular
the stop masses. A Higgs mass near 125 GeV therefore requires
relatively heavy stops, which in turn tends to require corrections to
the Lagrangian parameters that are larger than the tree--level values.

Quantifying the resulting finetuning is far from trivial,
however. Most analyses now define finetuning via the sensitivity
measures first introduced in \cite{Barbieri:1987fn}. Applying this to
the weak scale Higgs mass parameters, which determine the size of the
vacuum expectation values breaking the electroweak gauge symmetry, one
finds that finetuning increases quadratically with the supersymmetric
higgs(ino) mass parameter $\mu$ (at tree level), and with the soft
supersymmetry breaking masses of the stop squarks (at one--loop level)
and of the gluino (at two--loop level). Refs.
\cite{Feng:1999mn,Kitano:2006gv,weiler_n} therefore define natural
supersymmetry to
contain rather light higgsinos, perhaps somewhat heavier stop squarks
and still not very heavy gluinos. All other superparticles can be out
of reach of the LHC without leading to undue finetuning, at least as
defined in this manner.

We adopt this definition of natural supersymmetry in our
  analysis. It is based on an analysis of the Higgs potential at the
  electroweak scale, without assumptions about high--scale physics
  (e.g., boundary conditions for the soft terms); however, it makes
  the implicit assumption that weak--scale masses of the relevant
  superparticles are independent parameters. It is minimal in the
  sense of requiring the minimal number of relatively light particles
  needed to cancel the leading quadratically divergent radiative
  corrections to the mass of the Higgs boson in the SM. This feature
  allows a complete scan of the relevant parameter space using a
  cluster of a few thousand CPUs. This is why we focus on this
  ``minimal natural supersymmetry'' here.\footnote{The ``most
  natural'' spectrum does depend on how exactly finetuning is
  defined. For example, if the ``fundamental'' parameters are defined
  at a high scale where supersymmetry breaking is mediated to squarks
  and gluinos, the finetuning constraint on the gluino mass is often
  stronger than that on stop masses \cite{Casas:2014eca}. On the other
  hand, refs.\cite{tev_tune} argue that stop masses well above 1 TeV
  may well be natural once the {\em total} radiative corrections to
  the Higgs potential at the weak scale are considered, due to
  cancellations between different terms. Note that in these
    scenarios the weak--scale parameters are functions of -- often
    fewer -- ``fundamental'' parameters at the input scale, in which
    case the weak--scale parameters are {\em not} independent of each
    other.}

In particular, the first and second generation squarks are assumed to
be decoupled. This avoids constraints from supersymmetric ``flavor
excitation'' reactions like $q q' \rightarrow \tilde q \tilde q'$ as
well as $q g \rightarrow \tilde q \tilde g$, whose cross sections can
easily exceed those for $q \bar q \rightarrow \tilde t \tilde t^*,
\tilde g \tilde g$ if $m_{\tilde q} \sim m_{\tilde g}, m_{\tilde t}$.
At the same time, ``natural SUSY'' in this sense requires relatively
light higgsinos, third generation squarks at the TeV scale or
(preferably) below, and a gluino not far above $1$ TeV. ATLAS and CMS
have started to optimize searches for such scenarios and a relatively
large number of ``natural SUSY'' searches have by now been published.

However, as already noted, for a given mass third generation scalars
have much lower production cross sections than first generation
squarks. In addition, the cascade decays of third generation scalars
can be quite involved; in particular, top quarks in the final state
decay into three jets, or into a single $b-$jet plus a charged lepton
and a neutrino. These cascade decays tend to spread the signal over
several final states. These two effects imply that limits on third
generation sparticles are generally weaker than for first generation
squarks. So far, experimental limits have been only derived on
simplified natural SUSY scenarios involving at most three sparticles
and with simplifying assumptions on the decay modes
\cite{Aad:2015pfx,Chatrchyan:2013xna}. In this paper we instead
consider {\it realistic} natural SUSY scenarios which are consistent
with low energy limits as well as the observation of the 125 GeV Higgs
boson.

Phenomenological studies of natural SUSY scenarios at the LHC have
been performed in \cite{Buchmueller:2013exa, Baer:2012uy, Han:2013kga,
  Belanger:2015vwa,Kobakhidze:2015scd,Kowalska:2013ica}. However, these earlier papers either did not
investigate the whole parameter space of natural SUSY, or they did not
use the entire available set of LHC searches, recast for a complete
natural SUSY framework. 

In this work, we look at the phenomenology of fairly general natural
SUSY scenarios. We parameterize the spectrum of relevant
superparticles with six free parameters: the masses of $SU(2)$ singlet
(``right--handed'') and doublet (``left--handed'') stop squarks, the
higgsino mass parameter, the gluino mass, the trilinear stop sector
soft breaking parameter, and the ratio of vacuum expectation values of
the two Higgs doublets. We define these parameters directly at the
weak scale, without imposing any relations between them.  We randomly
sample this six dimensional parameter space with about 22,000 spectra,
all of which have the correct Higgs mass and the lightest neutralino
as the LSP, assuming flat priors. This set of model points covers a
large number of collider signatures at the LHC. We simulate all signal
processes and pass them to a fast detector simulation. These signal
events are then confronted with current ATLAS and CMS searches. We
consider the relevant searches for natural SUSY as well as inclusive
SUSY searches at the LHC. We show allowed and excluded regions of
parameter space. While we do not combine searches in a statistical
sense, we show that many searches contribute to the final limits.

The remainder of this article is organized as follows. In
Sect.~\ref{sec:natural_susy_setup} we define the natural SUSY scenario
we consider, and qualitatively discuss its collider signatures. In
Sect.~\ref{sec:numerical_analysis} we first discuss the numerical
tools employed for this study and then describe how we perform the
scan. Our numerical results are shown in
Sect.~\ref{sec:numerical_results}. We conclude in
Sect.~\ref{sec:summary}.

\section{Natural Supersymmetry Setup} 
\label{sec:natural_susy_setup}
 
The goal of ``natural'' SUSY models is to minimize finetuning while
accommodating a 125 GeV Higgs boson as well as the negative results of
searches for superparticles. The finetuning in question is associated
with the spontaneous breaking of the electroweak gauge symmetry.  In
the Minimal Supersymmetric extension of the Standard Model (MSSM) the
higgsino mass parameter $\mu$ is fixed by the minimization conditions
of the Higgs potential \cite{susy},
\beq \label{en1}
\mu^2 = \frac { m_{H_d}^2 - m_{H_u}^2\tan^2\beta } {\tan^2\beta - 1} -
\frac{1}{2} M_Z^2.
\eeq
Here $m^2_{H_{u,d}}$ are the squared soft supersymmetry breaking
masses of the Higgs doublet giving masses to up--type and down--type
quarks, respectively, $\tan\beta$ is the ratio of the vacuum
expectation values of $H_u^0$ and $H_d^0$, and $M_Z$ is the mass of
the $Z$ boson. In a ``natural'' theory each individual term on the
right--hand side of eq.(\ref{en1}) should be at most of order
$M_Z^2$. An immediate consequence is that $\mu$ should also be of
order $M_Z$, leading to relatively light higgsinos in the spectrum.

The minimization condition (\ref{en1}) holds for running parameters
defined at scale $Q_{\rm EW}$; a common choice is $Q_{\rm EW} =
\sqrt{m_{\tilde t_1} m_{\tilde t_2}}$ where $\tilde t_1, \tilde t_2$
are the two stop mass eigenstates, since this approximately minimizes
the leading radiative corrections to the Higgs potential
\cite{Gamberini:1989jw}. The soft breaking masses in eq.(\ref{en1})
are subject to radiative corrections. The leading one--loop
corrections involve stop squarks and scale with $m^2_{\tilde
  t_{1,2}}$. Keeping these corrections small thus indicates that stop
masses should be as small as possible \cite{weiler_n}. At two loop
order, gluino loops also renormalize the soft breaking Higgs masses;
keeping these corrections under control requires gluino masses not too
much above the TeV scale \cite{weiler_n}.

\begin{table*}[t!]
\begin{center}
\begin{tabular}{|c|c|c|}
\hline
Parameter & Description & Scanned range \\
\hline
$m_{\tilde Q_t}$& 3$^{\rm rd}$ generation $SU(2)$ doublet soft
breaking squark mass & $[0.1 \ {\rm TeV},\, 1.5 \ {\rm TeV}]$\\
$m_{\tilde t_R}$ & 3$^{\rm rd}$ generation $SU(2)$ singlet soft
breaking squark mass & $[0.1 \ {\rm TeV},\, 1.5 \ {\rm TeV}]$\\
$M_3$ & Gluino mass parameter & $[0.1 \ {\rm TeV},\, 3.0 \ {\rm TeV}]$\\
$A_{t}$ & Stop trilinear coupling & $[-3.0 \ {\rm TeV},\, 3.0 \ {\rm TeV}]$\\
$\mu$ & Higgsino mass parameter & $[0.1 \ {\rm TeV},\, 0.5\ {\rm TeV}]$\\
$\tan\beta$ & Ratio of vacuum expectation values & $[1, \, 20]$ \\
\hline
\end{tabular}
\end{center}
\caption{Variable input parameters of our natural SUSY scenario, and
  the range over which these parameters are scanned. In case of $\mu$
  we give the range of the absolute value; negative values of $\mu$
  are also sampled. Note that the ranges refer to the running
  $\overline{\rm DR}$ parameters, defined at scale $Q = 1$ TeV.} 
 \label{tab:softbreaking} 
\end{table*}

It should be recognized that these arguments are somewhat
qualitative. Obviously the upper bounds on higgsino, stop and gluino
masses depend on how much finetuning one is willing to
tolerate. Moreover, the precise definition of the ``fundamental''
parameters of the theory, including the energy scale at which they are
defined, also matters \cite{Casas:2014eca}. Here we follow the spirit
of ref.\cite{weiler_n} and define our ``minimal natural SUSY''
scenario to have higgsinos with masses below 500 GeV, third generation
scalar quarks with masses less than 1.5 TeV and gluinos with mass
below $3$ TeV. Given the upper bound on the stop masses, the light
CP--even Higgs boson $h$ can attain its observed mass near 125 GeV
only if the trilinear $|A_t|$ soft breaking term is quite large. The
variable input parameters defining our natural SUSY scenario are
listed in Table~\ref{tab:softbreaking}.

In addition we assume a common large mass for the first and second
generation squarks, $\tilde b_R$ squarks and all sleptons, which we
fix to $m^2_{\tilde f} = 1.5 \cdot 10^7$ GeV$^2$. This easily
satisfies constraints from the null results of SUSY searches from
ATLAS and CMS, avoids constraints from flavor changing neutral
currents and alleviates bounds from CP violating processes
\cite{Gabbiani:1996hi}. Note that we only consider scenarios with
$\tan\beta \leq 20$, so that $\tilde b$ loops are subdominant and
$\tilde b_R$ can be made heavy. ($m_{\tilde b_L} = m_{\tilde t_L}$ by
gauge invariance.) We decouple the electroweak gauginos as well,
setting $M_1 = M_2 = 3$ TeV.  Since the observed Higgs boson is
SM--like, we are working in the decoupling limit with a large mass of
the pseudoscalar Higgs, $m_A = 2.5$ TeV. The precise values of these
masses basically do not matter for our analysis, as long as these
particles are well beyond the range of LHC run--1. Since the bottom
trilinear soft breaking coupling $A_b$ has only little impact on the
phenomenology we set $A_b=0$ for simplicity. Since $m^2_{\tilde b_R}
\gg m^2_{\tilde b_L}$, $L-R$ mixing in the sbottom sector is
suppressed, in contrast to the stop sector, where the mass eigenstates
generally are mixtures of $\tilde t_L$ and $\tilde t_R$
squarks. Finally, we assume $R-$parity to be conserved. This implies
that the LSP, which is stable, must be electrically neutral; in the
context of our scenario this means that the lightest neutralino must
be the LSP.

In most cases making all sparticles heavy that are not involved in the
simple finetuning argument outlined above should be conservative, in
the sense that additional light superparticles lead to additional
production channels which might exclude a scenario that is otherwise
allowed. Also, light selectrons or smuons might be produced in cascade
decays, increasing the rate of multi--lepton events which are
generally more tightly constrained than purely hadronic events. There
are two exceptions to this, however. The constraints on the direct
production of $\tilde \tau$ leptons are still quite weak, and the
$\tau$ tagging efficiency is not large. Allowing the $\tilde \tau$
sleptons to be light would therefore probably not make a scenario
easier to exclude. On the other hand, as long as the Bino and Wino
masses are very large, $\tilde \tau$ sleptons would not be produced in
stop or gluino decays even if they were light, so allowing light
$\tilde \tau$'s would probably not change our conclusions.

The second, and potentially more worrisome, exception is the Bino. If
first and second generation squarks are heavy the direct Bino
production cross section is very small, so a light Bino would not
change the total SUSY production cross section very much. On the other
hand a light Bino would allow scenarios where the lighter
(higgsino--like) chargino $\tilde \chi_1^+$ is heavier than the
lighter stop $\tilde t_1$. If in addition $m_t + m_{\tilde \chi_1^0}$
is only slightly smaller than $m_{\tilde t_1}$, $\tilde t_1
\rightarrow t + \tilde \chi_1^0$ decays would have a large branching
ratio (since this would be the only allowed two--body decay mode of
$\tilde t_1$), but a rather poor signature (since $\tilde t_1$ would
then look like a top quark, which has a much larger production cross
section and thus contributes a formidable background). This would make
searches even for quite light stops difficult, although not impossible
\cite{loophole}. This loophole is not really available in our
scenario, since we have $m_{\tilde \chi_1^0} \simeq m_{\tilde
  \chi_2^0} \simeq m_{\tilde \chi_1^+} \simeq |\mu| < m_{\tilde t_1}$,
where the last inequality follows from our demand that the LSP be the
lightest neutralino; hence $m_{t} + m_{\tilde \chi_1^0} \simeq
m_{\tilde t_1}$ would imply that many $\tilde t_1$ would decay into $b
+ \tilde \chi_1^+$, which (for $m_{\tilde \chi_1^+} \simeq m_{\tilde
  \chi_1^0}$) looks quite different from a top quark. However, even in
a more general model utilizing this loophole requires some finetuning,
which arguably is against the spirit of natural supersymmetry.
Moreover, at least in standard cosmology a stable Bino--like LSP well
separated in mass from all other sparticles would have a much too high
cosmological relic density. Avoiding this would require additional
finetuning, e.g. by chosing $m_{\tilde \tau_1} \simeq m_{\tilde
  \chi_1^0}$ \cite{coann}. In contrast, our spectra, which feature a
higgsino--like, relatively light LSP, are cosmologically safe,
although in standard cosmology the bulk of the dark matter density
would have to be provided by some other particle, e.g. the axion
or/and the axino \cite{baer_axi}.

Finally, one might worry that even if the Bino is too heavy to
  be produced in stop or gluino decays the boundaries of the allowed
 regions might still depend on its mass, since it affects the mass
 splitting between the higgsino--like states, and hence the amount of
 visible energy produced in the decays of the heavier states. We show
 near the end of Sec.~IV that this is not the case.

Since in our scenario the only potentially accessible strongly
interacting superparticles are gluinos and third generation squarks,
the most important production channels are:
\beq\label{eq:production}
pp\rightarrow \tilde g \tilde g,\quad 
pp\rightarrow \tilde t_{1(2)}\tilde t_{1(2)}^*,\quad 
pp\rightarrow \tilde b_1 \tilde b_1^*,
\eeq
where all sparticle production processes can be accompanied by
additional initial and/or final state radiation. We have omitted the
production of higgsino pairs since the small splitting between the
higgsino mass eigenstates, typically ${\cal O}(1)$ GeV in our
scenario, make the $\tilde \chi_2^0$ and $\tilde \chi_1^\pm$ decay
products too soft for higgsino pair production to be detectable at the
LHC. On the other hand, the mass difference is sufficiently large for
these decays to be effectively prompt \cite{gunion}. Hence the
chargino and the heavier neutralino mass eigenstate can be treated as
missing energy, just like the lightest neutralino. However, the
production of a higgsino pair in association with a jet (monojet
signature) is negligible at the LHC since the production rate is too
small even for Run 2 of the LHC \cite{higgsino_mono}.

Depending on the ordering of the states in the spectrum, the decay
chains can be relatively complicated. If kinematically allowed, the
third generation squarks will dominantly decay via the strong
interaction into a gluino and a quark:
\beq \label{strongdec}
\tilde t_a\rightarrow t \tilde g\  (a=1,2), \quad 
\tilde b_1\rightarrow b \tilde g\,.
\eeq
The squarks can always undergo two--body decay via Yukawa interactions:
\bea \label{yukdec}
\tilde t_a &\rightarrow& t \tilde\chi_l^0 \ (l=1,2), \quad
b\tilde\chi_1^+ \, ,\\
\tilde b_1 &\rightarrow& b \tilde\chi_l^0 \ (l=1,2), \quad
t\tilde\chi_1^-\, .
\eea
Since the top Yukawa coupling is quite large, the branching ratios for
decays (\ref{yukdec}) will be sizable even if the strong decays
(\ref{strongdec}) are allowed. As shown above, all three higgsino
states effectively act as missing energy in our scenario.

In addition, the squarks can have purely bosonic decay modes:
\bea
\tilde t_a &\rightarrow& \tilde b_1 W^+\ (a = 1,2)\,, \quad \tilde b_1
\rightarrow \tilde t_1 W^- \,, \label{eq:heavy_stop_decay1} \\
\tilde t_2 &\rightarrow& \tilde t_1 Z,\quad 
\tilde t_2 \rightarrow \tilde t_1 h. \label{eq:heavy_stop_decay2}
\eea
Since in our scenario the two--body decay $\tilde t_1 \rightarrow b
\tilde \chi_1^+$ is practically always allowed, as noted above,
tree--level three-- or four--body decays, as well as the loop--induced
decay $\tilde t_1 \rightarrow c \tilde \chi_1^0$, which in general can
be quite important \cite{Hikasa:1987db}, do not play a role in our
scan. 

If kinematically allowed, gluinos decay via the tree--level two--body
modes
\beq \label{gl2}
\tilde g \rightarrow \tilde t_a \bar t,\quad \tilde t_a^* t \
(a=1,2)\, ,
\quad \tilde b_1 \bar b, \quad\tilde b_1^* b\,.
\eeq
If these decays are kinematically suppressed, tree--level three--body
decays via off--shell third generation squarks are possible:
\beq \label{gl3}
\tilde g \rightarrow b \bar b \tilde \chi^0_l, \quad t \bar t \tilde
\chi_0^l \ (l=1,2)\, \quad b \bar t \tilde \chi_1^+, \quad \bar b t
\tilde \chi_1^-\,.
\eeq
All gluino decay modes in eqs.(\ref{gl2}) and (\ref{gl3}) give rise to
a large $b-$jet multiplicity. For a given mass the gluino, being a
color octet fermion, has a much larger cross section than the scalar
color triplet third generation squarks. This indicates that searches
for final states containing $b-$jets and missing $E_T$ will play an
important role in probing our natural SUSY scenario. However, if all
squarks are much heavier than the gluino, the tree--level two--body
modes (\ref{gl2}) are closed, and the modes (\ref{gl3}) are strongly
suppressed by the off--shell squark propagator. In this case the
loop--induced gluino decays \cite{Baer:1990sc,Chalons:2015vja}
\beq \label{gll}
\tilde g \rightarrow g\tilde\chi_l^0 \ (l = 1,2)  
\eeq
become important.

\section{Numerical Analysis} 
\label{sec:numerical_analysis} 

In this section, we first discuss the numerical tools used in this
work. Then we describe the generation of model points, summarize
theoretical and low energy constraints and describe the framework for
testing the model points against LHC data. 
 
\subsection{Numerical Tools} 
\label{subsec:numerical_tools}

The masses and decay branching ratios for each model point in the scan
are calculated with {\tt SPheno3.3.2} \cite{Porod:2011nf}. For
benchmark points with a compressed spectrum\footnote{The precise
  definition is given in the next subsection}, a matched sample of
signal events including up to one additional parton (i.e. matching
parton--level $2 \rightarrow 2$ and $2 \rightarrow 3$ events) is
generated with {\tt Madgraph5.1.2} \cite{Alwall:2011uj} interfaced
with {\tt Pythia 6.4} \cite{Sjostrand:2006za} for showering and
hadronization. Otherwise, the entire event generation is handled by
{\tt Pythia 8.185} \cite{Sjostrand:2014zea}. We rescale the cross
sections with a flat $k$-factor of 1.5. The truth level MC events are
passed on to {\tt CheckMATE1.2.1} \cite{Drees:2013wra,webpage,manager} which
is based on the modified fast detector simulation {\tt Delphes 3.10}
\cite{deFavereau:2013fsa}. {\tt CheckMATE} tests if the model point is
excluded or not at $95\%$ confidence level by comparing with published
experimental searches at the LHC. Since {\tt CheckMATE} uses the
background estimates provided by the experiments as part of their
analyses, no background events had to be generated by us.

\subsection{Scan Procedure}

We have performed a multidimensional scan in the parameters of natural
SUSY. To that end, we have randomly generated sets of the free
parameters within the ranges shown in Table~\ref{tab:softbreaking},
assuming flat probability distributions. The lower bounds on the
masses reflect the results of searches at lower energy colliders. In
particular, searches at LEP exclude model points where a charged
superparticle is lighter than about 100 GeV. All other soft breaking
parameters are given constant large values, as described in
Sec.\ref{sec:natural_susy_setup}. The sign of $\mu$ is also chosen
randomly with equal a priori probability for positive and negative
$\mu$.

The values of these input parameters are passed on to the spectrum
generator {\tt SPheno}, which computes the on--shell masses from the
input $\overline{\rm DR}$ parameters. {\tt SPheno} also applies
theoretical and experimental constraints on the spectrum. All
benchmark points must have correct electroweak symmetry breaking, and
all sfermions must have positive squared masses. We also demand that
the lightest CP--even Higgs boson must have a mass $m_h=125\pm3$ GeV,
where the range is an estimate of the uncertainty of the
calculation. We discard model points where the LSP is not the lightest
neutralino. We require the mass difference between the lighter
chargino and the LSP to exceed 150 MeV, in which case $\tilde
\chi_1^\pm$ decays are prompt. The presence of long--lived heavy
charged particles in the event would constitute a good signature with
little SM background \cite{stable}.

We checked that the stop loop contribution to the electroweak $\rho$
parameter \cite{Drees:1990dx} is always within the experimental
limits, and we expect this to be true for other electroweak precision
observables as well, due to the decoupling property of supersymmetric
particles. An unpublished combined exclusion limit on the chargino
mass by all four LEP collaborations place a lower limit of 103.5
GeV. However, for very small mass differences, the limit becomes
weaker. We impose a lower mass limit of 100 GeV on the lightest
chargino eigenstate from data of the LEP2 run \cite{Abbiendi:2002vz,
  Abdallah:2003xe, Abbiendi:2003sc}. We do not explicitly apply any
Tevatron limits as a preselection of the benchmark points; recall,
however, that we only sample spectra with gluino mass above 100 GeV.

\begin{figure} 
\includegraphics[width=0.5\textwidth,scale=0.4]{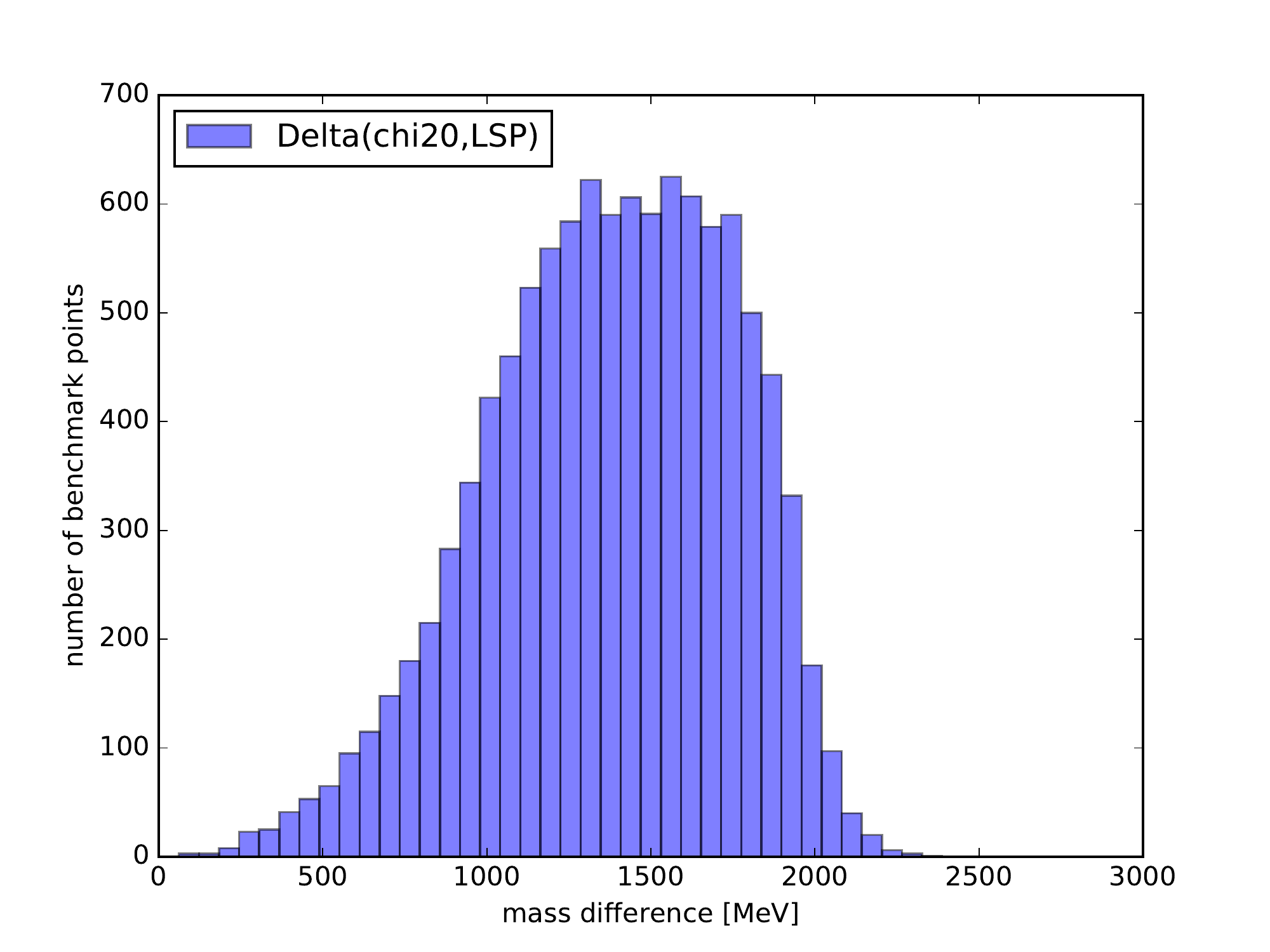} 
\caption{Distribution of the mass splitting between
  $\tilde\chi_2^0$ and $\tilde\chi_1^0$ for our model points
  satisfying the preselection constraints.} 
\label{fig:number_delta} 
\end{figure}

We have randomly generated about $22,000$ model points satisfying all
the preselection cuts. Since the higgsino mass eigenstates are nearly
mass degenerate, the NLSP is in general the lighter chargino mass
eigenstate or the second lightest neutralino. In
Fig.~\ref{fig:number_delta} we show the mass splitting between the LSP
and the second lightest neutralino. For a mass splitting larger than
the pion mass, the decay of the heavier neutralino is effectively
prompt. However, this figure also shows that the visible $\tilde \chi_2^0$
decay products will almost always be too soft to be observed at the
LHC. 

\begin{figure} 
\includegraphics[width=0.5\textwidth,scale=0.4]{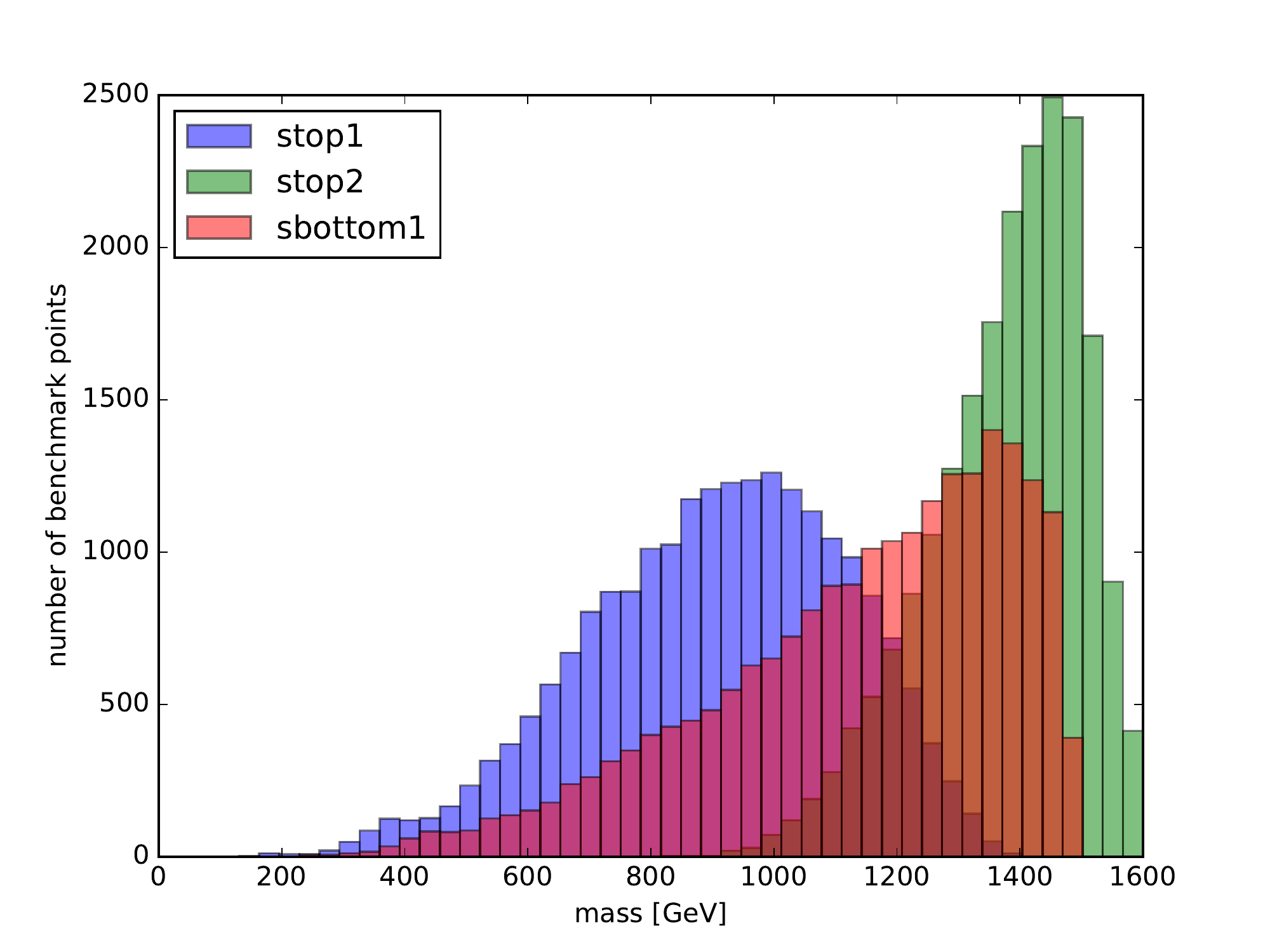} 
\caption{Distributions of the masses of the third generation squarks
  $\tilde t_1$, $\tilde t_2$ and $\tilde b_1$ for our model points
  satisfying the preselection constraints.} 
\label{fig:number_scalar} 
\end{figure}

In Fig.~\ref{fig:number_scalar} we show histograms of the scalar
particle masses. We see a significant mass splitting between the
lighter and the heavier stop mass eigenstates. For relatively light
$\tilde t_1$, a rather heavy $\tilde t_2$ is required to obtain a
sufficiently large value of $m_h$ in the MSSM; the smallest value of
$m_{\tilde t_2}$ among the 22,000 model points is just under 800
GeV. Since even within simplified models current lower bounds on third
generation sparticle masses do not exceed 700 GeV, $\tilde t_2$ pair
production by itself does not lead to significant constraints on our
model points. Thus we will be sensitive to the bosonic decays $\tilde
t_2$ as described in Eq.~(\ref{eq:heavy_stop_decay2}) mostly if
$\tilde t_2$ is produced in gluino decays. 
 
As expected, the mass of the lighter sbottom covers a large range. If
the lighter stop mass eigenstate is dominantly a $SU(2)$ doublet, a
light sbottom mass eigenstate with similar mass will also emerge in
the sparticle spectrum. Bosonic $\tilde b_1\rightarrow \tilde t_1
W^\pm$ decays, see Eq.~(\ref{eq:heavy_stop_decay1}), can be important
as long as $\tilde t_1$ has a significant $SU(2)$ doublet
component. This could lead to observable signals even if direct stop
pair production is not observable because the mass splitting to the
LSP is too small. Finally, recall that we fixed the $\tilde b_R$ mass
to a very large value, so $\tilde b_R$ production does not play any
role in our scan.

\begin{figure} 
\includegraphics[width=0.5\textwidth,scale=0.4]{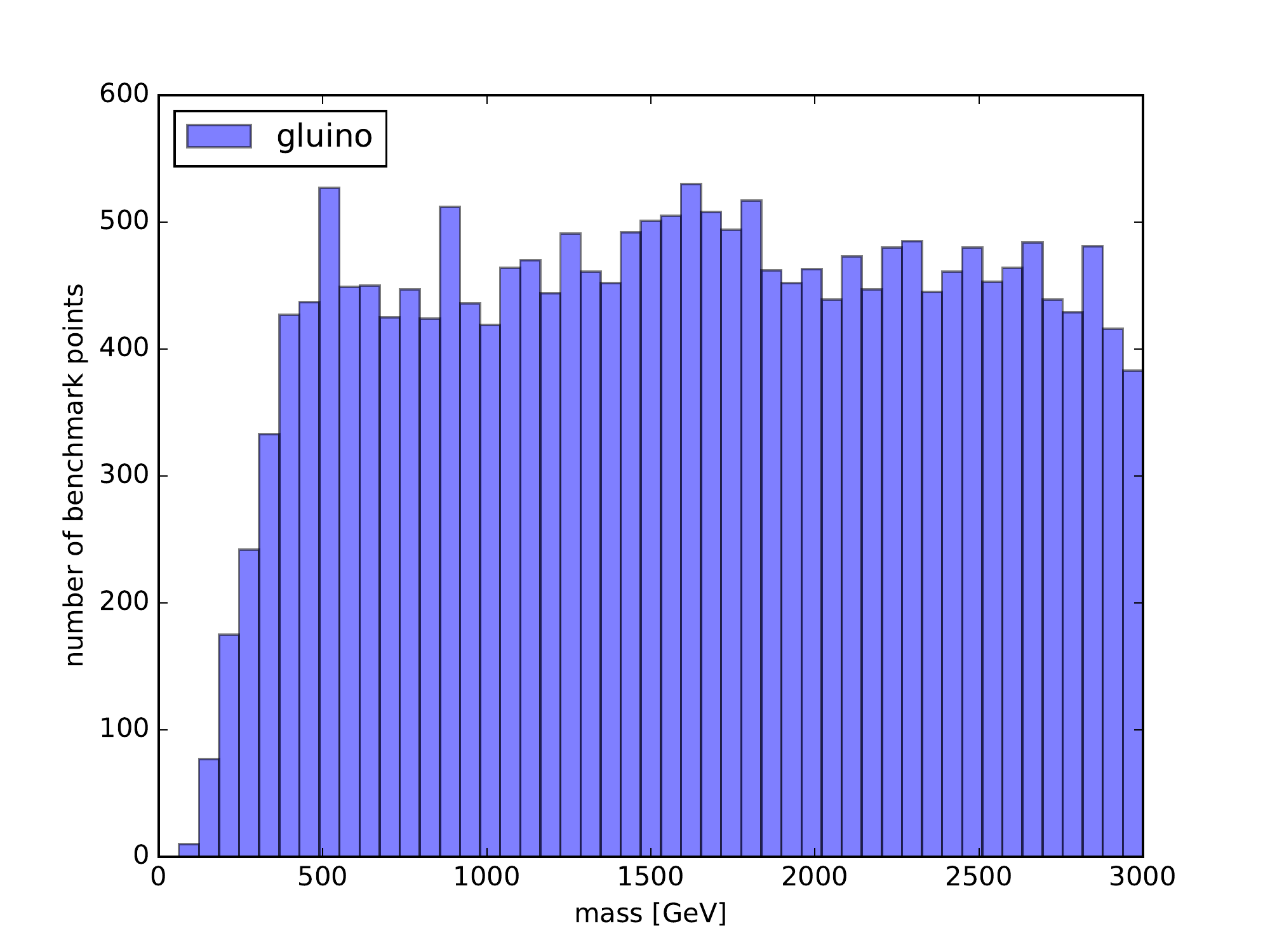} 
\caption{Distributions of the masses of the gluino for our model
  points satisfying the preselection constraints.} 
\label{fig:number_gluino} 
\end{figure}

Fig.~\ref{fig:number_gluino} shows the gluino mass distribution. It is
basically flat above 500 GeV; this is not unexpected, since the soft
breaking ($\overline{\rm DR}$) gluino mass is one of the input
parameters that is randomly sampled from a flat distribution. The
distribution falls off below 500 GeV since we discard points with
$m_{\tilde g} < m_{\tilde \chi_1^0} \simeq |\mu|$, and we require
$|\mu| < 500$ GeV. In simplified models gluino pair production
followed by gluino decay into third generation (s)quarks can be probed
by published LHC searches for gluino masses up to $1.3$ TeV. We also
sampled model points with significantly heavier gluinos since in
principle combinations of gluino and third generation squark pair
production might exclude model points where squark and gluino
production by itself satisfies all constraints. Moreover, we wanted
to have a statistically meaningful sample where effectively only the
third generation squarks could contribute to signatures at run 1 of
the LHC.

We saw in the last section that natural supersymmetry scenarios cover
a large number of final state topologies. The relative importance of
these topologies depends on the details of the particle spectrum,
e.~g. on the mass ordering and the mixing between the stop current
eigenstates. Fortunately, both LHC experiments ATLAS and CMS have
covered a large number of final states relevant for the production of
stops, sbottoms as well as gluinos. Moreover, both experiments have
performed powerful ``inclusive'' SUSY searches, targeting final states
with (generally untagged) jets and a large amount of missing $E_T$, in
some cases also requiring the presence of charged leptons. These
inclusive searches were used primarily to derive limits on the
parameter space of constrained supersymmetric models, but they can
also be sensitive to our natural SUSY scenario. By considering all these
searches together we expect to obtain improved limits on the parameter
space. 

The relevant searches implemented in {\tt CheckMATE} are listed in
Table~\ref{tab:lhc_searches}. The left column gives an identifier for
the given search; published results are identified by their arXiv
number, while results from conference proceedings are identified with
their ATLAS or CMS internal number. The second column of this Table
show the final state signature for which the given analysis is
optimized. The third column gives the total integrated luminosity used
in that analysis. More details about these twelve experimental
searches are given in Appendix~\ref{app:analyses}.

It should be mentioned that the preponderance of ATLAS searches is
simply due to the historical accident that {\tt CheckMATE} currently
has implemented many more ATLAS than CMS analyses. Generally ATLAS and
CMS searches for a given final state show similar sensitivity. Since
we do not statistically combine different searches, adding CMS
searches for the final states also searched for by ATLAS would not
change our results very much. We do include a couple of CMS searches
that do not have a close ATLAS equivalent.

\begin{table}
\begin{center}
\begin{tabular}{l|l|l}
Reference & Final State & $\mathcal{L}$ [fb$^{-1}$]\\
\hline
1308.2631 (ATLAS) \cite{Aad:2013ija} & 0$\ell$+2$b$ jets+\met & 20.1 \\
1403.4853 (ATLAS) \cite{Aad:2014qaa} & 2$\ell$+\met& 20.3 \\
1404.2500 (ATLAS) \cite{Aad:2014pda} & SS 2$\ell$ or 3$\ell$ & 20.3\\
1407.0583 (ATLAS) \cite{Aad:2014kra} & 1$\ell$+($b$) jets+\met & 20.0\\
1407.0608 (ATLAS) \cite{Aad:2014nra} & monojet+\met & 20.3\\
1303.2985 (CMS) \cite{Chatrchyan:2013lya} & $\alpha_{T}$+$b$ jets & 11.7\\
ATLAS-CONF-2012-104 \cite{ATLAS-CONF-2012-104} & 1$\ell$+$\geq$ 4
jets+\met & 5.8\\ 
ATLAS-CONF-2013-024 \cite{ATLAS-CONF-2013-024} & 0$\ell$+6 (2$b$)
jets+\met & 20.5\\ 
ATLAS-CONF-2013-047 \cite{ATLAS-CONF-2013-047} & 0$\ell$+2-6 jets+\met
& 20.3 \\ 
ATLAS-CONF-2013-061 \cite{ATLAS-CONF-2013-061} & 0-1$\ell$+$\geq$ 3$b$
jets+\met & 20.1\\ 
ATLAS-CONF-2013-062 \cite{ATLAS-CONF-2013-062} & 1-2$\ell$+3-6
jets+\met & 20.0\\ 
CMS-SUS-13-016 \cite{CMS-PAS-SUS13-016} & OS 2$\ell$+$\geq$3$b$ jets & 19.7
\end{tabular}
\end{center}
\caption{The experimental analyses used in our study. The *CONF*
  papers are only published as conference proceedings, the others are
  given by their arXiv number. The middle column denotes the final
  state for which the analysis is optimized, and the third column
  shows the total integrated luminosity employed in this analysis.}
\label{tab:lhc_searches} 
\end{table}

In order to predict the number of signal events for all signal regions
of the various analyses several simulation steps are needed. We first
generate $5,000$ truth level Monte Carlo (MC) events for each
benchmark point, including all processes given in
Eq.~(\ref{eq:production}). The corresponding total cross section
before cuts is also computed at this step; as noted above, we scale up
the leading order cross section by a universal ``$k-$factor'' of
$1.5$.

For strongly interacting sparticles with relatively small mass
splitting to the LSP, leading to rather soft visible decay products,
an accurate treatment of additional radiation is important; in the
extreme case of very small mass splitting this can give rise to a
monojet signature \cite{monojet}, which has been searched for in one
of the analyses we consider \cite{Aad:2014nra}. We therefore describe
the pair production of third generation squarks or gluinos with mass
less than 300 GeV above the LSP mass by matching event samples with
two and three partons in the final state, using the parton-jet MLM
matching algorithm described in Ref.~\cite{Mangano:2006rw}. The
numerical matching is performed with {\tt Madgraph} interfaced with
the shower generator {\tt Pythia6.4}, where we applied a $p_T$ sorted
parton shower and hadronization is switched on. We generate $50,000$
MadGraph events for each of these model points; this tenfold increase
overcompensates the fact that some events are removed in the matching
process. In order to reduce the required computational effort, the
production of superparticles with larger mass splitting to the LSP is
directly handled by {\tt Pythia8.185} without matching. In some cases,
{\tt Pythia} was not able to hadronize final states from sparticle
decays with very small splitting and we removed those model points.

\begin{figure} 
\includegraphics[width=0.5\textwidth,scale=0.4]{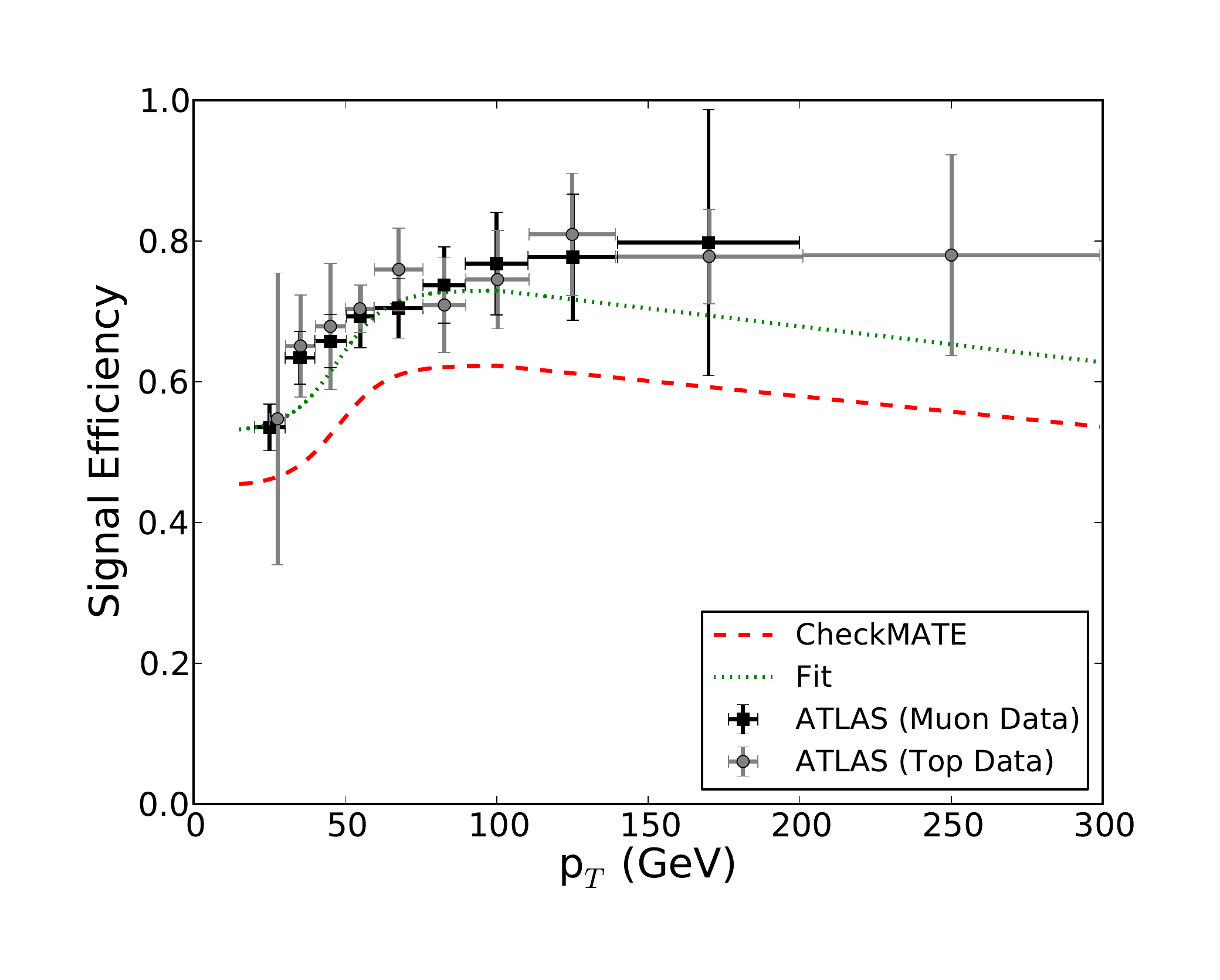} 
\caption{The data points show $b-$tagging efficiencies as determined
  by the ATLAS collaboration from two different analyses
  \cite{ATLAS-CONF-2012-097,ATLAS-CONF-2012-043}, for one specific
  working point. The dotted line is a fit to these data points, and
  the red dashed curve shows the actual {\tt CheckMATE}
  implementation, which has been scaled down by a factor 0.85 as
  described in the text.}
\label{fig:bjet_efficiency} 
\end{figure}

The truth level MC events are then further processed with the tool
{\tt CheckMATE}. It consists of a simulation of the detector response
with a modified {\tt Delphes} where the settings have been retuned to
mimic the responses of the ATLAS detector. In particular, an accurate
description of the $b-$tagging efficiency is of crucial importance
since, as shown in Table~\ref{tab:lhc_searches}, many third generation
searches rely on $b-$tagging. The tagging efficiency measured by ATLAS
and its implementation in {\tt CheckMATE} are shown in
Fig.~\ref{fig:bjet_efficiency} as a function of the $p_T$ of the
$b-$jet. Here the parameters of the tagging algorithm have been chosen
such that the overall $b-$tagging efficiency for $t \bar t$ events is
$70\ \%$.  Two different data sets have been used to derive the fit
shown in this figure, with different sensitivities at small and large
momenta \cite{ATLAS-CONF-2012-097, ATLAS-CONF-2012-043}. Moreover, the
overall normalization has been reduced by $15 \ \%$ in order to obtain
better agreement between the {\tt CheckMATE} implementation of ATLAS
and CMS new physics searches and the actual experimental results. This
could be due to the final states in new physics searches typically
being more complicated than the ones used to determine the $b-$tagging
efficiency shown in Fig.~\ref{fig:bjet_efficiency}.

The reconstructed detector level objects, as well as the total cross
sections computed earlier, are then passed on to the analysis module
of {\tt CheckMATE}. All studies listed in Table~\ref{tab:lhc_searches}
have been implemented in {\tt CheckMATE} and have been carefully
validated against the results published by the experiments. {\tt
  CheckMATE} typically reproduces the total cut efficiency given by
the experiments with an accuracy of $10\%$ or better. More details on
the validation of all implemented experimental searches can be found
in the {\tt CheckMATE} manual and web page
\cite{Drees:2013wra,webpage}.

We check each model point against all the analyses given in
Table~\ref{tab:lhc_searches}.  Note that each of these analyses
defines several ``signal regions'', defined by sets of kinematic
cuts. Out of these many regions, {\tt CheckMATE} finds the one with
the largest {\em expected} exclusion potential; this is computed from
the background determined by the experimenters and its error, as well
as the signal cross section times cut efficiency for this particular
signal region, and is {\em independent} of the actually observed
number of events in this signal region. Finally, for the signal region
selected in this manner, {\tt CheckMATE} compares the sum of the
background and the predicted signal with the actual experimental
observation and determines if the model point is excluded at the
$95\%$ C.L., using the so--called CL$_S$ method \cite{Read:2002hq}. 
More specifically, it computes the parameter
\beq \label{r}
r \equiv \frac{S-1.96\cdot\Delta S} {S_{\rm exp.}^{95}}\,, 
\eeq
where $S$ is the number of signal events, $\Delta S$ denotes its
theoretical uncertainty, and and $S_{\rm exp.}^{95}$ is the
experimentally determined 95$\%$ confidence level limit on the
signal. We only include the error from the limited statistics of our
Monte Carlo simulation, i.e. $\Delta S = \sqrt{S}$. The actual (as
opposed to expected) value of $r$ is only computed for the ``optimal''
signal region defined above, in order to avoid spurious exclusions due
to downward fluctuations in the data; since the analyses we employ
define well over a hundred signal regions, we expect several $2
\sigma$ fluctuations to have occurred in these data. In order to keep
the statistical analysis simple and transparent, {\tt CheckMATE} does
not statistically combine signal regions of a particular analysis, nor
does it combine different analyses. {\tt CheckMATE} considers a model
to be excluded at 95$\%$ c.l.  level if $r$ defined in eq.(\ref{r})
exceeds $1$. According to this strict definition, Monte Carlo
fluctuations would decide whether scenarios with $r \approx 1$ are
considered excluded or allowed. We therefore increase the event sample
to $50,000$ whenever the original assessment of $r$ gave a value
between $2/3$ and $3/2$. Moreover, we conservatively consider a model
point to be (definitely) excluded only if our final estimate gives $r
\ge 1.5$, while points with $r \le 2/3$ are considered (definitely)
allowed. Model points with $2/3 < r < 3/2$ are thus
indeterminate. This can be considered to be a (probably rather
conservative) simple method to include theoretical uncertainties on
the predicted signal strengths.

\section{Limits on Gluino and Third Generation Scalar Masses}  
\label{sec:numerical_results}  

We are now ready to present numerical results of our scan. We first
make some general remarks and then show the distribution of allowed
and excluded points in the planes spanned by two of the three most
important model parameters, which are the masses of the gluino, of the
lighter stop, and of the LSP. We also delineate completely excluded as
well as completely allowed regions of parameter space. Finally, we
discuss the properties of model points which evade current collider
searches.

As previously described, we have randomly generated $22,000$ model
points. In the majority of these points all superparticles are beyond
the reach of LHC run 1, and thus these models points are still
allowed. However, about $25\ \%$ of all model points are ruled out by
the experimental searches we consider ($r > 1.5$) and
another $6.6 \ \%$ are indeterminate ($2/3 < r < 3/2$).

\begin{table*}[t!]
\begin{center}
\begin{tabular}{|l|l|c|c|c|c|c|}
\hline
Experiment & Final State & \multicolumn{4}{|c|}{Best Sensitivity} &
Excludes \\
 & & all & excluded & ambiguous & allowed &  \\
\hline
ATLAS \cite{ATLAS-CONF-2013-061} & 0-1$\ell$+$\geq$ 3$b$ jets+\met &
0.22 & 0.37 & 0.56 & 0.13 & 0.57 \\ 
ATLAS \cite{ATLAS-CONF-2013-047} & 0$\ell$+2-6 jets+\met & 0.37 & 0.25
& 0.056 & 0.44 & 0.69 \\
CMS \cite{Chatrchyan:2013lya} & $\alpha_{T}$+$b$ jets & 0.088 & 0.11 &
0.14 & 0.075 & 0.66 \\
ATLAS \cite{ATLAS-CONF-2013-024} & 0$\ell$+6 (2$b$) jets+\met & 0.044
& 0.12 & 0.041 & 0.016 & 0.58 \\ 
ATLAS \cite{Aad:2014kra} & 1$\ell$+($b$) jets+\met & 0.14 & 0.078 &
0.10 & 0.16 & 0.45 \\
ATLAS \cite{Aad:2014nra} & monojet+\met & 0.013 & 0.042 & 0.018 &
0.002 & 0.23 \\
ATLAS \cite{Aad:2013ija} & 0$\ell$+2$b$ jets+\met & 0.10 & 0.019 &
0.085 & 0.13 & 0.051 \\
ATLAS \cite{ATLAS-CONF-2013-062} & 1-2$\ell$+3-6 jets+\met & 0.024 &
0.002 & 0.001 & 0.034 & 0.50 \\ 
ATLAS \cite{Aad:2014pda} & SS 2$\ell$ or 3$\ell$ & 0.0 & 0.0 & 0.0 &
0.0 & 0.070 \\
ATLAS \cite{ATLAS-CONF-2012-104} & 1$\ell$+$\geq$ 4 jets+\met & 0.0 &
0.0 & 0.0 & 0.0 & 0.12 \\
CMS \cite{CMS-PAS-SUS13-016} & OS 2$\ell$+$\geq$3$b$ jets & 0.0 & 0.0
& 0.0 & 0.0 & 0.043 \\
ATLAS \cite{Aad:2014qaa} & 2$\ell$+\met& 0.0 & 0.0 & 0.0 & 0.0 & 0.0
\\
\hline
\end{tabular}
\end{center}
\caption{Statistical information about the sensitivity of the twelve
  analyses with respect to our natural SUSY scenarios. The first
  column gives the name of the experiment and the references. The
  corresponding final state is given in the second columns. Columns
  three through six give the fraction of model points for which this
  analysis is the {\em most sensitive} one; column 3 is for the entire
  set of model points, while the next three columns only include
  model points with $r > 1.5, \ 2/3 < r < 3/2$ and
  $r < 2/3$, respectively. The last column gives the
  fraction of all excluded model points that are (also) excluded by
  this particular analysis, i.e. where this analysis has $r
  > 1.5$.}
\label{tab:lhc_sensitivityI} 
\end{table*}

Table~\ref{tab:lhc_sensitivityI} compiles statistics about the twelve
considered analyses. Columns 3 through 6 give the fractions of model
points for which a given analysis is the most sensitive one. This is
shown with respect to the entire set of model points (col. 3), as well
as specifying to clearly excluded (col. 4), ambiguous (col. 5) and
clearly allowed model points (col. 6). 

We see that the ATLAS search \cite{ATLAS-CONF-2013-061} for final
states with missing transverse momentum and at least three $b$ jets
performs best among the clearly excluded and ambiguous points. This is
not surprising since we expect a large number of $b$ jets from direct
gluino pair production with subsequent decay into third generation
sparticles. The ATLAS inclusive multijet plus missing $E_T$ search
with a charged lepton veto \cite{ATLAS-CONF-2013-047} also plays an
important role in constraining natural SUSY even though the study is
not optimized for this scenario. In fact, we see that this search
offers the best sensitivity for clearly allowed points. This indicates
that in future inclusive SUSY searches might be more important in
further constraining the currently still allowed parameter space of
natural SUSY than dedicated searches for third generation
squarks. This is related to our upper bound $|\mu| < 500$ GeV, which
ensures that model points where all strongly interacting sparticles
are well beyond current sensitivity limits will have large mass
splitting to the LSP. This implies good sensitivity for the inclusive
search, without having to pay the price in efficiency that is required
by multiple $b-$tags.

While either of these two analyses performs best in nearly two third
of the excluded model points, a total of six further searches
sometimes have the best sensitivity. This shows the importance of
including a large set of experimental searches when constraining the
parameter space even of our relatively simple implementation of
natural supersymmetry. In particular, the inclusive CMS $\alpha_T$
analysis \cite{Chatrchyan:2013lya} classifying events according to
their $b$ jet multiplicity performs quite well despite the fact that
the data sample only corresponds to an integrated luminosity of 11.7
inverse femtobarn. The ATLAS single lepton search \cite{Aad:2014kra}
also does a reasonably good job. In our case decays of the stop into
$\tilde t_1\rightarrow t \tilde\chi_1^0$, or three--body gluino decays
including at least one top quark in the final state, will result in
many events with one isolated lepton.

On the other hand, we see that searches that require two or more
charged leptons never have the best sensitivity to our model
points. The rate for two lepton final states is heavily suppressed by
the leptonic branching ratio. This might change, however, if we
allowed sleptons to be relatively light, which would e.g. be required
if loops with supersymmetric particles were to explain the $\sim 3
\sigma$ deviation in the anomalous magnetic moment of the muon. It
should be noted that the multivariate analysis targeting the decay
mode $\tilde t_1\rightarrow t \tilde\chi_1^0$ in \cite{Aad:2014qaa} is
not implemented in this work, since {\tt CheckMATE} currently only
allows to implement cut--based analyses; the other signal regions in
\cite{Aad:2014qaa} are optimized for stop decays into charginos with
leptonic chargino decays and thus dilepton final states from $\tilde
t_1\rightarrow t \tilde\chi_1^0$ decays are frequently missed. Same
sign tops, leading to events with two same sign leptons, can be
produced in the decays of the gluino. An analogous signature can also
arise from gluino mediated sbottom production with the subsequent
decay $\tilde b_1\rightarrow t \tilde \chi_1^\pm$. Events with same
sign leptons have a very small SM background. However, again due to
the small cumulative branching ratio the SS dilepton search does not
perform as well as the other searches.

The last column in Table~\ref{tab:lhc_sensitivityI} shows the fraction
of clearly excluded model points (i.e. points with $r > 1.5$) that are
excluded by this analysis, i.e. where the best signal region of a
given analysis has $r > 1.5$. The entry in this column is obviously
larger than that in the second column, which only counts the fraction
of model points for which this particular analysis performs best. The
sum of the entries in this column is significantly larger than 1,
showing that many disallowed points are in fact excluded by several
independent analyses. Even the searches for final states with two or
more leptons, which never offer the best sensitivity as we just saw,
do exclude some of our points, i.e. they help to increase the
confidence level with which these points can be excluded; the only
exception is the ATLAS search of ref.\cite{Aad:2014qaa}, which does
not exclude any of our model points.

Now we want to discuss where in the overall parameter space 
the experimentally excluded as well as the clearly allowed model
points are situated. To this end we show in Figs.~5 to 8 the planes
spanned by two of the most relevant masses, which are the masses of
the gluino, of the lighter stop eigenstate, and of the LSP. In these
figures clearly excluded points (with $r > 1.5$) are marked
in red, and clearly allowed points (with $r < 2/3$) are
shown in green. Ambiguous points are not shown at all, in order to
better illustrate the separation between allowed and excluded regions
of parameter space.

\begin{figure} 
\includegraphics[width=0.5\textwidth,scale=0.4]{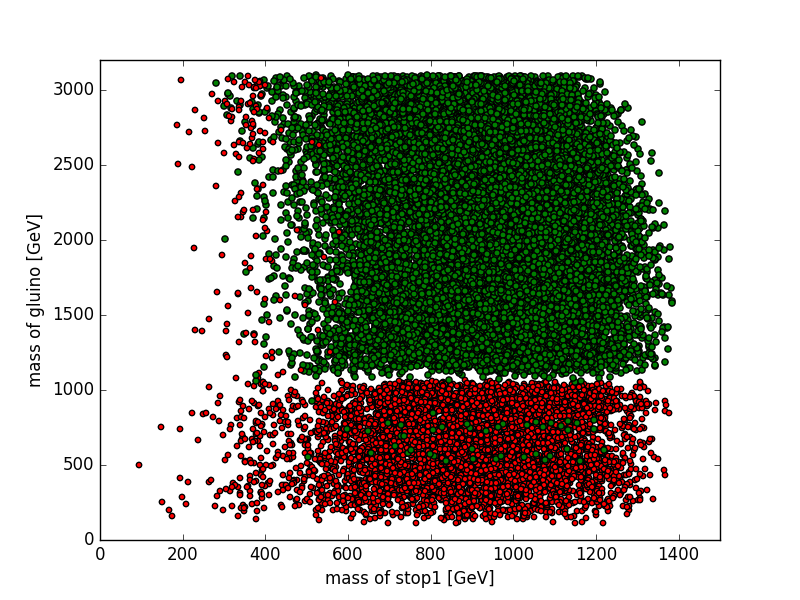} 
\caption{Models points in the stop and gluino mass plane. 
Model points clearly passing all constraints are shown in green, while
clearly excluded points are shown in red.}
\label{fig:meff_mgluino_r} 
\end{figure} 

Fig.~\ref{fig:meff_mgluino_r} shows the $\tilde t_1$ and $\tilde g$
mass plane. Not surprisingly, points where both these masses are large
cannot be excluded, simply because the total cross section for the
production of superparticles was too small at run 1 of the
LHC. Specifically, we see that no scenarios that satisfy $m_{\tilde
  t_1} > 580$ GeV and $m_{\tilde g} > 1070$ GeV are excluded.
Similarly, all points with $m_{\tilde t_1} > 660$ GeV and $m_{\tilde
  g} > 1180$ GeV are clearly allowed. The gap at $m_{\tilde g} \simeq
1100$ GeV is due to the ambiguous points. The fact that a clear gap
appears indicates that our definitions of allowed and excluded points
are indeed conservative, i.e. few if any points get labeled allowed
or excluded just due to Monte Carlo fluctuation; the latter only shift
points between the clearly allowed and ambiguous categories, as well
as between the clearly excluded and ambiguous categories.

While Fig.~\ref{fig:meff_mgluino_r} allows to define a region of the
$m_{\tilde g}, m_{\tilde t_1}$ plane where all points are allowed, and
a slightly larger region where no points are excluded, finding a
region of this plane where all points are excluded is not so easy. For
example, we see that some points with $m_{\tilde g}$ well above $1.2$
TeV and $m_{\tilde t_1} < 600$ GeV are excluded, but other points with
similar combination of gluino and lighter stop mass are allowed.
Similarly, there are a few allowed points with gluino mass well below
1 TeV.

\begin{figure} 
\includegraphics[width=0.5\textwidth,scale=0.4]{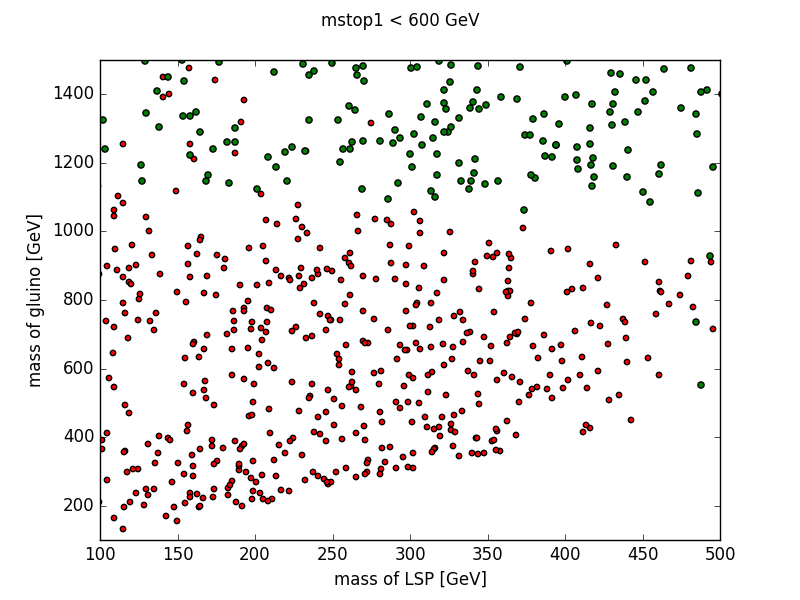} 
\caption{Model points with the lighter stop mass less than 600 GeV in
  the LSP and gluino mass plane. The notation is as in
  Fig.~\ref{fig:meff_mgluino_r}.}  
\label{fig:mlsp_mgluino_mst1_LT_600} 
\end{figure} 

The reason for this intermingling of allowed and excluded points is
that Fig.~\ref{fig:meff_mgluino_r} does not distinguish between
different values of the LSP mass. The mass gap between the LSP and the
directly produced strongly interacting superparticles largely
determines the amount of visible energy, and of missing transverse
momentum, in the event. In Fig.~\ref{fig:mlsp_mgluino_mst1_LT_600} we
therefore present the allowed (excluded) model points in green (red)
in the LSP and gluino mass plane while demanding that the lighter stop
mass eigenstate is lighter than 600 GeV. We see that now no model
points with gluinos masses less than 1000 GeV and LSP mass below 480
GeV are clearly allowed. Only a few model points with $m_{\tilde g} <
1$ TeV are allowed, which have LSP mass near the upper limit of our
scan, leading to a relatively small amount of visible energy in the
events. Similarly, all points with $m_{\tilde t_1} < 600$ GeV,
$m_{\tilde g} < 950$ GeV and $m_{\tilde \chi_1^0} < 400$ GeV are
clearly excluded.

A certain number of model points with heavy gluino, $m_{\tilde
  g}\ge1100$ GeV, are also excluded. It is clear from our discussion
of Fig.~\ref{fig:meff_mgluino_r} that in these cases the exclusion is
mostly due to direct stop and, perhaps, sbottom pair
production. However, since for equal masses the $\tilde t_1$ and
$\tilde b_1$ pair production cross sections are far smaller than the
$\tilde g$ pair production cross section, direct squark pair
production only excludes a relatively small region of parameter
space. In particular, we see from
Fig.~\ref{fig:mlsp_mgluino_mst1_LT_600} that even for relatively light
$\tilde t_1$, $m_{\tilde t_1} < 600$ GeV, no points with $m_{\tilde g}
> 1100$ GeV and $m_{\tilde \chi_1^0} > 300$ GeV are clearly excluded.

\begin{figure} 
\includegraphics[width=0.5\textwidth,scale=0.4]{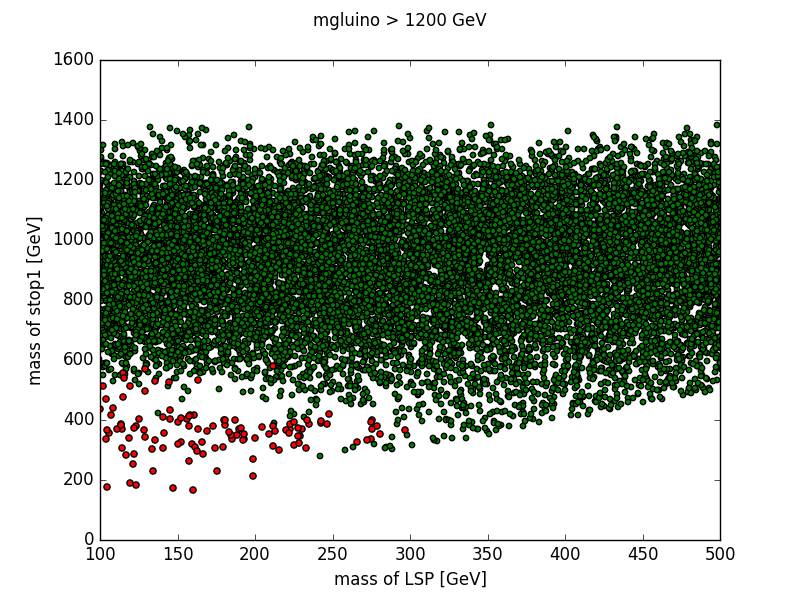} 
\caption{Models points with the gluino mass greater than 1200 GeV in
  the LSP and stop mass plane. The notation is as in
  Fig.~\ref{fig:meff_mgluino_r}.}
\label{fig:mlsp_mst1_mgl_GT_1200} 
\end{figure}

The limited scope of searches for direct pair production of third
generation squarks is further illustrated in
Fig.~\ref{fig:mlsp_mst1_mgl_GT_1200}, which shows all points with
$m_{\tilde g} > 1.2$ TeV in the $\tilde t_1$ and LSP mass plane. We see
again that no point with heavy gluino and $m_{\tilde t_1} > 600$ GeV
is currently excluded. Even for LSP masses near the lower limit
allowed by LEP searches, there are allowed points with $m_{\tilde t_1}
\lsim 450$ GeV, and ambiguous points with $m_{\tilde t_1} \lsim 350$
GeV. Only the region $m_{\tilde t_1} < 300$ GeV, $m_{\tilde \chi_1^0}
< 210$ GeV is completely excluded by these searches.

\begin{figure} 
\includegraphics[width=0.5\textwidth,scale=0.4]{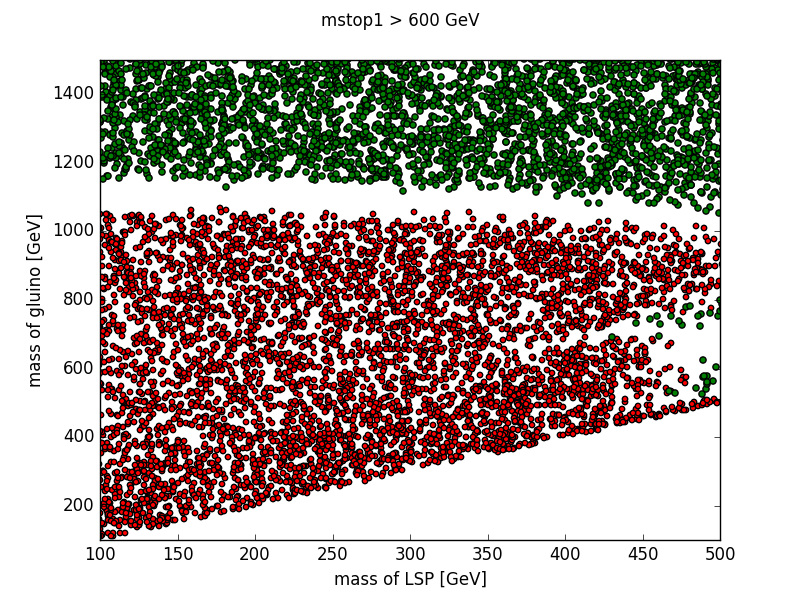} 
\caption{Models points with $m_{\tilde t_1} > 600$ GeV in the LSP and
  gluino mass plane. The notation is as in Fig.~\ref{fig:meff_mgluino_r}.}
\label{fig:mlsp_mgluino_mst1_GT_600} 
\end{figure}

Fig.~\ref{fig:mlsp_mgluino_mst1_GT_600} shows the plane spanned by the
gluino and LSP masses, and only includes points where squarks are too
heavy for direct squark pair production to exclude the scenario, i.e. 
$m_{\tilde t_1} \ge 600$ GeV. The limit is thus essentially determined by
the production of gluino pairs. Recall that in natural SUSY, gluinos
quite often decay into third generation squarks; in particular, final
states containing at least one $t \bar t$ pair are quite common also
in many of these scenarios.

We observe a clear separation between an entirely allowed and a mostly
excluded region. Again, the gap arises because we do not show
ambiguous model points which are neither clearly excluded nor clearly
allowed. The lower edge of the allowed region is reduced by about
$100$ GeV as the LSP mass increases from its lower bound near $100$
GeV to the maximal value near $500$ GeV considered in our definition
of natural supersymmetry; evidently the increased LSP mass reduces the
visible energy, and hence the efficiency of the cuts of the most
sensitive analyses. However, even for $m_{\tilde \chi^0_1}$ near its
upper bound the mass splitting to the gluino exceeds 500 GeV in this
large clearly allowed region, so we are not yet dealing with a
compressed spectrum.

This is different for the small island of allowed model points near
the right boundary of Fig.~\ref{fig:mlsp_mgluino_mst1_GT_600}, with
LSP mass above 430 GeV and gluino mass below 800 GeV. Fixing the LSP
mass near 500 GeV, we see that reducing the gluino mass below $1000$
GeV in many cases reduces the signal after cuts, i.e. the increase of
the total gluino pair production cross section is over--compensated by
the reduction of the cut efficiencies of the most sensitive
analyses. Somewhat surprisingly this allowed region does not extend
all the way down to the line $m_{\tilde \chi_1^0} = m_{\tilde g}$,
which are excluded by monojet searches in ref.\cite{Aad:2014nra} even
for the highest LSP masses in our scan. As a result, the region where
no points are excluded is quite small; for instance, no points with
$m_{\tilde t_1} > 600$ GeV, $m_{\tilde g} \in [600, \, 760]$ GeV and
$m_{\tilde \chi_1^0} > 470$ GeV can be excluded. However, even in this
narrow range of gluino masses, there are allowed points with LSP mass
down to about 430 GeV, and ambiguous points up to the highest LSP mass
in our scan. The reason is that the gluino branching ratios still
depend on the masses of third generation squarks, even if the latter
are too heavy for their pair production to contribute significantly to
the total SUSY cross section.

The various excluded, not excluded, and clearly allowed regions are
summarized in Table~\ref{tab:regions1}. Here an ``excluded (allowed)
region'' is a region in parameter space in which all model points are
excluded (allowed), while a ``not excluded'' region is a region where
no model points are excluded; the latter regions include ambiguous
points. 

\begin{center}
\begin{table}[h]
\begin{tabular}{|l|l|}

\hline
Type & Boundaries \\
\hline
Allowed & $m_{\tilde t_1} > 660$ GeV and $m_{\tilde g} > 1180$ GeV \\
 &        $m_{\tilde g} > 1150$ GeV and $m_{\tilde \chi_1^0} > 370$
 GeV \\
\hline
& $m_{\tilde t_1} > 580$ GeV and $m_{\tilde g} > 1070$
GeV \\
Not  & $m_{\tilde g} > 1060$ GeV and $m_{\tilde \chi_1^0} >
300$ GeV \\ excluded
 & $m_{\tilde t_1} > 550$ GeV, $m_{\tilde \chi_1^0} > 470$ GeV \\ & \ and
 $m_{\tilde g} \in [600 \ {\rm GeV},\, 760 \ {\rm GeV}]$\\
\hline
 & $m_{\tilde t_1} < 230$ GeV or $m_{\tilde g} < 440$ GeV \\
Excluded & $m_{\tilde g} < 990$ GeV and $m_{\tilde \chi_1^0} < 340$ GeV \\
 & $m_{\tilde g} < 1040$ GeV and $m_{\tilde \chi_1^0} < 200$
GeV \\ &
$m_{\tilde t_1} < 300$ GeV and $m_{\tilde \chi_1^0} < 210$ GeV \\
\hline

\end{tabular}
\label{tab:regions1}
\caption{List of allowed, not excluded and excluded regions. In the
  allowed regions, all model points have $r < 2/3$; in the
  excluded regions, all model points have $r > 1.5$; and in
  the not excluded regions, all model points have $r <
  1.5$}.
\end{table}
\end{center}

Note that there are points that are neither in one of the ``excluded''
nor in one of the ``not excluded'' regions listed in
Table~\ref{tab:regions1}. The reason is that this table defines
regions only based on the values of three parameters: the masses of
the gluino, of the lighter stop, and of the LSP. While these are the
most important parameters deciding whether a model point is excluded
or not, they are not the only ones. For example, the mass of the
lighter sbottom is also relevant. In our scan we always have
$m_{\tilde b_1} > m_{\tilde t_1}$, but the mass difference is
typically quite small if $\tilde t_1$ is mostly an $SU(2)$ doublet,
which requires $m_{\tilde Q_t} < m_{\tilde t_R}$. The presence of
$\tilde b_1$ only slightly above $\tilde t_1$ obviously increases the
total squark pair production cross section. Moreover, $\tilde b_1$ and
$\tilde t_1$ pair production often yield essentially the same final
state, if $\tilde t_1 \rightarrow t \tilde \chi_{1,2}^0$ and $\tilde
b_1 \rightarrow t \tilde \chi_1^-$; if kinematically allowed, these
are typically the most important decay modes if $\tilde t_1$ is mostly
an $SU(2)$ doublet. In contrast, a mostly $SU(2)$ singlet $\tilde t_1$
prefers to decay into $b \tilde \chi_1^+$ even if the decays into $t
\tilde \chi_{1,2}^0$ are kinematically allowed. The reason is that for
the relevant case of higgsino--like lighter chargino, the $\tilde t_R
b \tilde \chi_1^\pm$ coupling is proportional to the top Yukawa
coupling, while the $\tilde t_L b \tilde \chi_1^\pm$ coupling is
proportional to the much smaller bottom Yukawa coupling. This
difference in $\tilde t_1$ decay modes also affects the $r$
value of a model point.

The upshot of this discussion is that some combinations of $m_{\tilde
  g}, \, m_{\tilde t_1}$ and $m_{\tilde \chi_1^0}$ can be allowed,
excluded or ambiguous depending on the values of the other
parameters. However, Table~\ref{tab:regions1} shows that over much of
the parameter space these three parameters suffice to determine the
fate of a model point.

In some cases strong dependence of the cut efficiency on kinematic
quantities, together with fluctuations of the numbers of events
actually observed in certain search regions, also leads to quite large
differences in $r$ between points that are quite close in parameter
space. For instance, we found a pair of model points, p74 and p11081
in our scan, with quite similar spectra and decay branching ratios,
yet their $r$ values differ by a factor of two. In both cases the
gluino is so heavy that its production can be ignored, while both
$\tilde t_1$ and $\tilde b_1$ lie at or below 600 GeV, while the LSP
mass is relatively light. As explained above, both $\tilde t_1$ and
$\tilde b_1$ then decay predominantly into a top quark and a
higgsino--like neutralino or chargino; recall that the two heavier
higgsino--like states effectively behave the same way as the LSP, as
far as LHC signatures are concerned. Point p11081 has about $9\%$
heavier squarks, and about $40\%$ heavier higgsinos, such that the
energy of the top in the rest mass of the decaying $\tilde t_1$ is
nearly the same in both cases. Due to the larger squark masses this
scenario has nearly two times smaller squark production cross section
than p74, yet it yields a two times larger value of $r$; as a result,
p11081 is clearly excluded, while p74 is ambiguous.

We found that the small kinematic differences between the two
scenarios lead to significantly different efficiencies in the ATLAS
search for a hadronically decaying top pair plus missing $E_T$
\cite{ATLAS-CONF-2013-024}. As a result, the signal region expected to
be most sensitive to p11081 is from this search, whereas for p74 it is
from \cite{Aad:2014kra}, which searches for final states with one
lepton, two jets and missing $E_T$. Moreover, the relevant signal
region in \cite{ATLAS-CONF-2013-024} contains fewer events than
expected from backgrounds, while the relevant signal region in
\cite{Aad:2014kra} contains somewhat more events than predicted in the
background--only hypothesis; these differences are likely due to
fluctuations. As a result, the actual $r$ value is higher
than expected for p11081, but lower than expected in
p74.\footnote{The LHC experiments use the $S_{95}$ method \cite{s95} of
  setting limits. This ensures that an under--fluctuation by more than
  two standard deviations, which formally rules out the SM at the
  $95\%$ c.l. (without ``look elsewhere'' effect), does not exclude
  all scenarios where the expected number of events is larger than in
  the SM. This is essential, since given the large number of signal
  regions included in the analysis, it is highly likely that some
  fluctuate down by more than two standard deviations. However, such a
  downward fluctuation still does increase the $r$ value even in the
  $S_{95}$ method, as indeed it should.} This, together with the
strong dependence of the cut sensitivity of the relevant analysis in
\cite{ATLAS-CONF-2013-024}, leads to the counter--intuitive outcome
that only the heavier spectrum is excluded.

Finally, it is worth noting that, at least within our definition of
natural SUSY, we can derive absolute lower bounds of 440 GeV and 230
GeV on the mass of the gluino and the lighter stop squark,
respectively. These hold for all choices of the other parameters, in
particular also for very compressed spectra. As noted above, monojet
searches play an important role in deriving these absolute lower
bounds. 

\begin{center}
\begin{table}[h]
\begin{tabular}{|l|l|}

\hline
Type & Boundaries \\
\hline
Allowed & $m_{\tilde t_1} > 630$ GeV and $m_{\tilde g} > 1150$ GeV \\
 &        $m_{\tilde g} > 1100$ GeV and $m_{\tilde \chi_1^0} > 320$
 GeV \\
\hline
 & $m_{\tilde t_1} < 260$ GeV or $m_{\tilde g} < 480$ GeV \\
Excluded & $m_{\tilde g} < 1040$ GeV and $m_{\tilde \chi_1^0} < 340$ GeV \\
 & $m_{\tilde g} < 1070$ GeV and $m_{\tilde \chi_1^0} < 200$
GeV \\ &
$m_{\tilde t_1} < 390$ GeV and $m_{\tilde \chi_1^0} < 230$ GeV \\
\hline

\end{tabular}
\label{tab:r1}
\caption{List of allowed and excluded regions, where we now demand
  that in the allowed regions, all model points have $r <
  1$, while in excluded regions, all model points have $r >
  1$. There are no ambiguous points in this case.}
\end{table}
\end{center}

Recall that we call a model point (clearly) excluded only if $r_{\rm
  max} > 1.5$. This helped to separate excluded and allowed regions in
the figures discussed in this Section. Nevertheless this may be overly
conservative, since {\tt CheckMATE} already incorporates the
statistical Monte Carlo error of the simulation in the calculation of
$r$. In Table~\textrm{V} we therefore list ``allowed'' regions
where all model points satisfy $r < 1$ and ``excluded'' regions where
all model points have $r > 1$. Since there are no ambiguous points in
this definition, the ``not excluded'' regions listed separately in
Table~\ref{tab:regions1} are then identical to the ``allowed''
regions, and are therefore not listed separately in
Table~\textrm{V}.

Comparing Table~\textrm{V} with Table~\ref{tab:regions1} we see that
both the allowed and the excluded regions have become larger, since
the requirements defining these regions have become weaker. However,
the allowed regions in Table~\textrm{V} are still smaller than the
corresponding ``not excluded'' regions in Table~\ref{tab:regions1},
since the latter allow points with $r_{\rm max}$ up to $1.5$, while in
the former all points have to satisfy $r < 1$. In particular, the
island of compressed spectra discussed in
Fig.~\ref{fig:mlsp_mgluino_mst1_GT_600} and listed as third ``not
excluded'' region in Table~\ref{tab:regions1} does not appear in
Table~\textrm{V}. We saw in Fig.~\ref{fig:mlsp_mgluino_mst1_GT_600}
that some points in this region have $r < 2/3$; these points obviously
also satisfy the requirement $r < 1$ used to define an allowed model
point in Table~\textrm{V}. However, there are also model points
throughout this island with $1 < r < 1.5$, which are now counted as
excluded, making it impossible to define a contiguous allowed region
defined in terms of only the gluino, lighter stop and LSP masses in
this case. On the other hand, the other allowed and excluded regions
in Table~\textrm{V} do not differ too much from the corresponding
regions in Table~\ref{tab:regions1}. In particular, the absolute lower
bounds on the $\tilde t_1$ and gluino masses have increased by only
about $10\%$.

Finally, we want to investigate the impact of the decoupled sparticle
spectrum on our results. Here, we want to have a closer look at the
electroweak gaugino sector.  As discussed in
Section~\ref{sec:natural_susy_setup}, we have fixed the bino and wino
mass parameters to rather large values, $M_1=M_2=3$ TeV. As a result,
the mass splitting between the second lightest neutralino or the
chargino and the LSP (see Fig.~\ref{fig:number_delta}) is too
small to have an measurable effect on the LHC phenomenology.  However,
if the values of the gaugino mass parameters are significantly
lowered, the mixing between the $U(1)$ as well as $SU(2)$ gauginos and
the higgsinos will become larger and the mass gap between the second
lightest neutralino or the lighter chargino mass eigenstate and the
LSP will be widened. Hence, the decay products of the chargino and the
second lightest neutralino might become energetic enough to be
detected at the LHC. 

\begin{figure} 
\includegraphics[width=0.5\textwidth,scale=0.4]{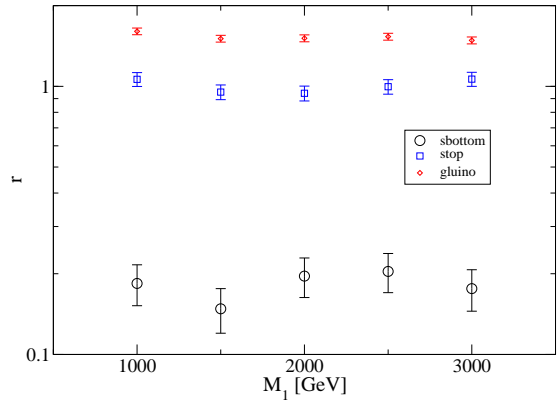} 
\caption{The value of $r$ a function of the soft breaking parameter $M_1$
  for the lighter stop, sbottom and gluino pair production. The other
  parameters are kept fixed. The results for stop and sbottom pair
  production have been computed with the same spectrum (but they probe
  different signal regions), while the results for gluino production
  are for a different model point, as explained in the text. The
  errors shown are due to Monte Carlo statistics.}
\label{fig:r_variation} 
\end{figure}

We checked whether this is the case by varying the Bino mass
  parameter $M_1$ while keeping all other parameters the same. We then
  computed the optimal value of $r$ from gluino, stop and sbottom
  production independently, rather than summing over all three
  modes. This probes the $M_1$ dependence of three different signal
  regions.  The results are depicted in Fig.~\ref{fig:r_variation}. We
  used the same model point for the stop and sbottom pair production,
  with $m_{\tilde t_1} \simeq 380$ GeV, $m_{\tilde b_1} \simeq 475$
  GeV and $\mu = 265$ GeV. Recall that our input parameters are
  $\overline{\rm DR}$ parameters; the physical squark masses do
  therefore vary slightly when $M_1$ is changed, but the variation is
  well below 1\%. We used a different scenario to probe the $M_1$
  dependence of the gluino pair production signal, with $\mu = 135$
  GeV, $m_{\tilde g} \simeq 1085$ GeV, $m_{\tilde t_1} \simeq 900$ GeV
  and heavy $\tilde b_1$ yielding $r \simeq 1.5$, since a possible
  $M_1$ dependence would be most important for us for scenarios with
  $r \simeq 1$, i.e. near the boundaries of the allowed and excluded
  regions.  In all three cases we kept $M_2$ fixed to 3 TeV. It is
  clear from the plot that $r$ is constant within the error induced by
  the Monte Carlo statistics. We conclude that the exact values of
  $M_1$ and $M_2$ are not relevant for our results as long as $|M_1|$,
  $|M_2|\gg |\mu|$.

We conclude this Section with some comments on model points which are
difficult to observe with run-1 SUSY searches. The simplest (and most
obvious) reason for this is that the total production cross section
before cuts becomes very small for heavy SUSY particles. Many of these
model points will be tested by run-2 of the LHC, which is currently
under way.

Some model points are kinematically accessible but still missed by all
searches. We encountered one reason for this already: a relatively
compressed spectrum, with small mass difference between the strongly
interacting superparticles and the LSP, greatly reduces the cut
efficiencies for most analyses. In the most extreme case one has to
fall back on monojet searches, where the signal only occurs at ${\cal
  O}(\alpha_S^3)$ and suffers from a large irreducible background from
$Z +$jet production.

Another class of difficult model points satisfies $m_{\tilde t_1}\sim
m_{\tilde\chi_1^0}+ m_t$, with $\tilde t_1 \rightarrow t \tilde
\chi^0_1$ being the dominant decay mode. In this case the LSPs in the
final state often have small momenta, so that the event resembles a $t
\bar t$ event. This leads to well--known holes in the ATLAS and CMS
exclusion plots in the $\tilde t_1$ and LSP mass planes. This has been
discussed in Sect.~\ref{sec:natural_susy_setup}, where we also pointed
out that this scenario can only be approximated in our definition
of natural supersymmetry: if the phase space for $\tilde t_1
\rightarrow t \tilde \chi_1^0$ becomes too small, $\tilde t_1
\rightarrow b \tilde \chi_1^+$ decays will take over, which are {\em
  always} allowed if $\tilde \chi_1^0$ is higgsino-like with
$m_{\tilde \chi_1^0} \simeq m_{\tilde t_1} - m_t$. One can try to
suppress the branching ratio into the chargino mode by making $\tilde
t_1$ mostly an $SU(2)$ doublet and chosing a small value of
$\tan\beta$. However, the latter cannot be too small, since we insist
on reproducing the observed mass of the Higgs boson. Moreover, a light
doublet--like $\tilde t_1$ implies that $\tilde b_1$ is also
relatively light, with mass typically 40 to 100 GeV above $m_{\tilde
  t_1}$. Hence the top quarks from $\tilde b_1$ decay will have
significant energy in the $\tilde b_1$ rest frame, so that $\tilde
b_1$ pair production can be distinguished from top pair production. As
a result, the smallest clearly allowed (ambiguous) $\tilde t_1$ mass
for model points with $m_{\tilde \chi_1^0} < m_{\tilde t_1} - m_t -
40$ GeV and branching ratio $B(\tilde t_1 \rightarrow t \tilde
\chi_{1,2}^0) > 0.5$ is about $390 \ (310)$ GeV (with $m_{\tilde b_1}
= 470 \ (430)$ GeV).  Furthermore, all model points where $\tilde t_1$
is lighter than $450$ GeV and can decay into top plus neutralino have
$r \geq 0.4$, and should thus be testable in future LHC runs.
 

\section{Summary and Conclusions}
\label{sec:summary} 

In this work, we have considered natural SUSY scenarios characterized
by light higgsinos with $|\mu|\le500$ GeV, relatively light third
generation $SU(2)$ doublet squarks and singlet stops, $m_{\tilde
  t_{1(2)}},m_{\tilde b_1}\le1.5$ TeV, and gluinos with mass
$m_{\tilde g}\le3$ TeV in order to preserve the electroweak
hierarchy. The first and second generation squarks, all sleptons as
well as the EW gauginos are decoupled; this avoids direct search
limits from ATLAS and CMS and suppresses FCNC and CP violating
processes. Since the observed Higgs couplings are consistent with the
SM predictions, we work in the decoupling limit where all additional
Higgs bosons predicted by the MSSM are also very heavy.

We have randomly generated 22000 natural SUSY model points in the six
dimensional parameter space: $m_{\tilde Q_t}$, $m_{\tilde t_R}$,
$\mu$, $M_3$, $A_t$ and $\tan\beta$ assuming flat priors. We demanded
a SM like Higgs boson with $m_h=125\pm3$ GeV and a neutralino LSP. In
this setup, the higgsino mass eigenstates are always almost mass
degenerate so that their decay products are too soft to be observed,
making it essentially impossible to probe direct higgsino pair
production at the LHC. On the other hand, stops, sbottoms as well as
gluinos can be copiously produced at the LHC. Novel decay signatures
such as heavier stop decays into $Z$ and $h$ or sbottom decays into
$W$ bosons emerge but these decay modes cannot be probed with current
LHC data since $m_{\tilde t_2} > 800$ GeV is required in order to
obtain a sufficiently heavy SM like Higgs and hence the heavier stop
could not produced at an observable rate during LHC run--1.

We have generated signal events for each model point and have passed
the event files to {\tt CheckMATE} which provides a framework to test
a model against a large number of current ATLAS and CMS 8 TeV searches
for beyond the SM physics, in particular, SUSY searches. We included
the results of searches optimized for simplified natural SUSY
scenarios, of direct squark and gluino searches and of inclusive SUSY
searches in our scan. All these searches have been implemented and
fully validated in {\tt CheckMATE}. We have found that nearly all the
searches we include indeed exclude some model points, and eight
different searches provide the best sensitivity in some region of
parameter space. This shows that considering a large number of
different searches is indeed necessary in order to determine whether a
model point is allowed or not.

The main results of our analysis are summarized in
Tables~\ref{tab:regions1} and \textrm{V}, which delineate allowed and
excluded regions in parameter space. In particular, we found that all
scenarios where either $m_{\tilde t_1} < 230$ GeV or $m_{\tilde g} <
440$ GeV are clearly excluded, irrespective of the values of the other
parameters. On the other hand, all model points with $m_{\tilde t_1} >
660$ GeV and $m_{\tilde g} > 1180$ GeV are currently still clearly
allowed. Here we call a model point ``clearly allowed'' only if its
predicted signal in the signal region which is expected to be most
sensitive to this point is at least $1.5$ times the nominal $95\%$
c.l. upper bound, while ``clearly allowed'' points have the a
predicted signal in this ``optimal'' signal region which is at least a
factor $1.5$ below the nominal bound. This serves to avoid overlap of
allowed and excluded regions due to Monte Carlo fluctuations. Note
also that we did not include any theoretical uncertainty of our
prediction beyond the statistical error of our Monte Carlo
simulation. This factor of $1.5$ can thus also be interpreted as a
(rather conservative) estimate of the additional theory
uncertainty. In many cases it is sufficient to specify the masses of
the lighter stop, the gluino, and the LSP in order to decide whether a
parameter point is excluded or allowed, but in some cases the values
of the other parameters are also important. In particular, there are
significant differences between points with doublet--like or
singlet--like lighter stop, not least because a doublet--like light
stop also implies a rather light sbottom. Moreover, we found some
cases where small differences in parameters can lead to large
differences in the ratio of the expected signal in the most sensitive
search region to its upper bound.

Overall we find that a large part of the parameter space of our
definition of natural supersymmetry is still allowed. Much of this
parameter space can be explored by run--2 of the LHC, which just
started. We forecast that inclusive SUSY searches will play an
increasingly prominent role in exploring the remaining parameter
space, while dedicated searches for final states containing top or
bottom quarks will play a lesser role than in the analysis of run--1
data. However, most likely again many analyses, and an even larger
number of signal regions, will have to be combined in order to
comprehensively probe the remaining parameter space. We look forward
to the results of the ongoing run.

 
\begin{acknowledgments} 
We thank Sabine Kraml for discussions. 
The work by M.D. was partially supported by the BMBF Theorieverbund. 
  The work of J.S.K. was partially supported by the MINECO,
  Spain, under contract FPA2013-44773-P; Consolider-Ingenio CPAN
  CSD2007-00042 and the Spanish MINECO Centro de excelencia Severo
  Ochoa Program under grant SEV-2012-0249. J.S.K. would like to
  thank Bonn university for support and hospitality while part of this
  manuscript was prepared.
\end{acknowledgments} 

\appendix

\begin{appendices}

\section{Analyses}\label{app:analyses}

Here we give brief descriptions of the analyses we used in our
scan. We focus on those aspects that are relevant for our definition
of natural SUSY, although our model points are also tested against
signal regions that are optimized for decay chains that cannot be
realized in our set--up.

\subsection{1308.2631 (ATLAS)}

This analysis \cite{Aad:2013ija} concentrates on signatures with two
$b-$jets and missing transverse momentum. It was optimized for sbottom
pair production followed by $\tilde b_1\rightarrow b
\tilde\chi_1^0$. A similar final state arises from stop pair
production followed by $\tilde t_1\rightarrow \tilde\chi_1^+ b$. The
search has two signal regions targeting scenarios with a large mass
splitting between the squark and the LSP or with a compressed
spectrum.  The former signal region demands large transverse momentum
and two $b-$jets and the latter requires a leading non $b$ jet
recoiling against the squark pair system with two $b$ tagged jets and
large transverse momentum.

\subsection{1403.4853 (ATLAS)}

This search for direct stop pair production in final states with two
leptons and large missing transverse momentum \cite{Aad:2014qaa}
targets scenarios with $\tilde t\rightarrow\tilde\chi_1^\pm b$ with
$\Delta m(\tilde\chi_1^\pm,\tilde\chi_1^0)\ge m_W$, or $\tilde
t\rightarrow t\tilde\chi_1^0$ with an on (off)-shell top quark. $t\bar
t$ and $W^+W^-$ production are the main background processes to this
search and the stransverse mass $m_{T2}$ observable
\cite{Lester:1999tx} can be used to suppress these backgrounds very
efficiently. The signal regions targeting $\tilde t_1\rightarrow
b\chi_1^\pm$ are divided according to jet multiplicity and $m_{T2}$,
whereas one signal region explicitly requires 2 $b-$jets. The
on--shell $\tilde t_1\rightarrow t \tilde\chi_1^0$ mode is addressed
via a multivariate method and is not implemented in {\tt CheckMATE}.

\subsection{1404.2500 (ATLAS)}

This analysis \cite{Aad:2014pda} considers final states containing two
same sign leptons or at least three leptons. This search was optimized
for gluino mediated stop production, $\tilde g\rightarrow t_1\tilde t$
with $\tilde t_1\rightarrow t\tilde\chi_1^0$. Here, one can expect up
to four leptons in the final state. The selection requirements of the
five signal regions differ in the number of $b-$tagged jets, jet
multiplicity, missing transverse momentum cut, threshold of the
effective mass, and the transverse mass computed from the $p_T$ of the
hardest lepton and the missing $p_T$.

\subsection{1407.0583 (ATLAS)}

This analysis \cite{Aad:2014kra} is designed to search for final
states containing one lepton, a minimum of two jets and large
transverse missing momentum. The study contains fifteen signal regions
targeting a large number of stop pair production scenarios,
with subsequent decay modes such as $\tilde t_1\rightarrow
t\tilde\chi_1^0$, $\tilde t_1\rightarrow b\tilde\chi_1^\pm$, $\tilde
t_1\rightarrow b f f^\prime\tilde\chi_1^0$, $\tilde t_1\rightarrow b
W\tilde \chi_1^0$, and non--symmetric decay modes such as $\tilde
t_1\rightarrow t \tilde \chi_1^0, \tilde t_1^* \rightarrow
\bar b\tilde\chi_1^-$. All signal regions include a veto on a second
lepton. The signal regions optimized for $\tilde t_1\rightarrow t
\tilde \chi_1^0$ decays use shape information of the large missing
transverse momentum and transverse mass distribution. If the mass
difference between the stop and neutralino is very large, the top
quark can be boosted and large--cone jets are used. The signal regions
targeting $\tilde t_1\rightarrow b \tilde \chi_1^\pm$ decays require
different kinematic cuts on the leptons, ($b-$)jets, missing
transverse momentum, transverse mass, asymmetric stransverse mass and
$b-$jet multiplicity or vetoes on isolated tracks and hadronic taus.

\subsection{1407.0608 (ATLAS)}

In scenarios where $\tilde t_1$ has small mass splitting to the LSP
the stop decay mode $\tilde t_1\rightarrow b\tilde\chi_1^0 W$ can be
kinematically closed, while the four--body decay $\tilde
t_1\rightarrow \ell\nu_\ell b\tilde \chi_1^0$ is strongly suppressed
because it is a third order process which is very sensitive to phase
space. Thus the loop--induced decay $\tilde t_1\rightarrow c\tilde
\chi_1^0$ can be the dominant decay mode \cite{Hikasa:1987db}. This
analysis \cite{Aad:2014nra} is optimized for searches for stop pair
production with $\tilde t_1\rightarrow c \tilde\chi_1^0$. The study
has defined two classes of signal regions. Both sets have the same
preselection cuts which require a hard jet, large missing transverse
momentum and a lepton veto. The first class of signal regions targets
scenarios with a very small mass splitting between the stop and the
neutralino LSP and thus the charm jets are too soft to be
reconstructible. The selection cuts thus isolate monojet events. The
second set of signal regions considers non--degenerate scenarios and
exploits a dedicated charm tagging algorithm. The signal regions are
further divided by the applied cuts on the momentum of the leading jet
and on the missing transverse momentum.

\subsection{1303.2985 (CMS)}

This analysis is designed to be sensitive to hadronic final states
with missing transverse energy using the variable $\alpha_T$
\cite{Chatrchyan:2013lya}. The sensitivity of the search is improved
by categorizing events according to the multiplicities of $b-$tagged
and other jets. The signal regions span a wide range of cuts on
the scalar sum of transverse energies of all the jets. Hence, the
search is sensitive to a large number of third generation simplified
models such as gluino mediated stop and sbottom production and direct
stop and/or sbottom production followed by their hadronic decay.

\subsection{ATLAS-CONF-2012-104}

This search targets final states with at least four hard jets, missing
transverse momentum and one lepton, and uses an integrated luminosity
of only 5.8 fb$^{-1}$ \cite{ATLAS-CONF-2012-104}. The study has two
non--overlapping search regions corresponding to an electron and a
muon channel. The event selection is mainly based on the transverse
mass of the lepton and missing transverse momentum, as well as on the
inclusive mass defined as the scalar sum of the transverse momenta of
the lepton, the jets and the missing transverse momentum. The cuts are
designed to efficiently suppress the dominant $t\bar t$ and $W/Z$+jet
backgrounds.  The search results were interpreted as limits on the
parameter space of minimal supergravity scenarios. In particular, the
search is motivated by scenarios where a left handed squark dominantly
decays to a light chargino. However, the search is also sensitive to
natural SUSY scenarios, such as gluino production with subsequent
decays into $t\bar b\tilde\chi_1^-$ or into stop and top with
semileptonic decay of the top and hadronic stop decay.

\subsection{ATLAS-CONF-2013-024}

This analysis searches for direct production of the top squark with
subsequent decay into top plus LSP \cite{ATLAS-CONF-2013-024}. It
concentrates on the purely hadronic decay mode of the top quark and
thus the all--hadronic stop search demands six or more jets while at
least two jets are tagged as $b-$jets. Each event is required to be
consistent with containing two top quarks and thus two 3--jet systems
must each have an invariant mass consistent with the top quark mass. The
signal definition includes a lepton as well as a tau veto and a
considerable amount of transverse missing momentum is required. The
search is divided into three signal regions with increasing minimum
missing transverse momentum cuts.

\subsection{ATLAS-CONF-2013-047}

This analysis \cite{ATLAS-CONF-2013-047} is designed to look for heavy
squark and gluino production in final states with high momentum jets,
large missing transverse momentum and no leptons, using a total
integrated luminosity of 20.3 fb$^{-1}$. The null results are
interpreted as limits on simplified models in the gluino (squark) and
neutralino LSP mass plane as well as in mSUGRA/CMSSM parameter
space. Since the search aims for heavy squark and gluino production
modes, $m_{\rm eff}$ is a powerful observable to separate the signal
from the SM background. $m_{\rm eff}$ is defined as the scalar sum of
the transverse momenta of the jets in the final state plus the missing
transverse momentum. The large number of signal regions with differing
jet multiplicity and kinematic requirements allow to target a broad
range of squark and gluino models from short to long cascade
decays. As no $b-$jet veto is applied, the search is also sensitive to
natural SUSY models as long as high $p_T$ jets and large missing
transverse momentum are expected in the final state.

\subsection{ATLAS-CONF-2013-061}

This multi $b-$jets study \cite{ATLAS-CONF-2013-061} aims at final
states with four or more jets, at least three of which originate from
$b-$quarks, and large missing transverse momentum. It uses an
integrated luminosity of 20.1 fb$^{-1}$. The signal regions are
defined via the number of charged leptons (zero or $\geq 1$), the jet
multiplicity, as well as different kinematic requirements on jet
momentum, missing transverse momentum and the effective mass. In
\cite{ATLAS-CONF-2013-061} the search results are interpreted in the
context of simplified natural SUSY scenarios and in the context of
mSUGRA/CMSSM scenarios, but this search is also very powerful in
constraining natural SUSY models since gluino pair production there
frequently leads to final states with four $b-$quarks. Several signal
regions target gluino decays into $\tilde g\rightarrow \tilde b b$ or
$\tilde g\rightarrow \tilde t t$, while others consider scenarios with
gluinos decaying via off--shell third generation squarks.

\subsection{ATLAS-CONF-2013-062}

This analysis \cite{ATLAS-CONF-2013-062} focuses on searches for squarks
and gluinos in final states with isolated (soft) leptons, jets and
missing transverse momentum, using a data set corresponding to an
integrated luminosity of 20 fb$^{-1}$. Limits are derived on
simplified gluino and stop pair production scenarios as well on the
mSUGRA/CMSSM model. This search is divided into five classes of signal
regions which are based on the inclusive hard single lepton channel, a
soft single lepton channel optimized for compressed spectra, the soft
dimuon channel addressing the mUED model and soft single lepton signal
regions with one or two $b-$jets targeting stop pair production for
small and moderate mass splitting between the stop and the neutralino
LSP, respectively.

\subsection{CMS-SUS-13-016 (CMS)}

This analysis \cite{CMS-PAS-SUS13-016} searches for superparticle
production in events with two opposite sign leptons, a large number of
jets, $b-$tagged jets, and large missing transverse energy. This
search is designed to search for gluino pair production with $\tilde
g\rightarrow t\bar t \tilde\chi_1^0$. It only has one signal region
which demands at least five jets, three of which are $b-$tagged, and
large missing transverse momentum.
\end{appendices}

\bibliographystyle{h-physrev}

\begin{thebibliography}{}

\end{thebibliography}


\begin{thebibliography}{99} 
 
\bibitem{susy} 
M. Drees, R.M. Godbole and P. Roy, ``Theory and Phenomenology of 
Sparticles'', World Scientific, Singapore (2004); 
H. Baer and X.R. Tata, ``Weak scale supersymmetry: From superfields to 
scattering events'', Cambridge University Press (2006). 
 
\bibitem{Ibanez:1991pr} 
  L.~E.~Ib\'a\~nez and G.~G.~Ross, 
  Nucl.\ Phys.\  B {\bf 368} (1992) 3. 
  
\bibitem{Kim:2014eva}
  J.~S.~Kim, K.~Rolbiecki, K.~Sakurai and J.~Tattersall,
  JHEP {\bf 1412} (2014) 010
  [arXiv:1406.0858 [hep-ph]].

\bibitem{Allanach:2014lca}
  B.~Allanach, S.~Biswas, S.~Mondal and M.~Mitra,
  Phys.\ Rev.\ D {\bf 91} (2015) 1,  011702
  [arXiv:1408.5439 [hep-ph]].
  
\bibitem{Chamseddine:1982jx}
  A.~H.~Chamseddine, R.~L.~Arnowitt and P.~Nath,
  Phys.\ Rev.\ Lett.\  {\bf 49} (1982) 970.
  
\bibitem{Alwall:2008va}
  J.~Alwall, M.~P.~Le, M.~Lisanti and J.~G.~Wacker,
  Phys.\ Rev.\ D {\bf 79} (2009) 015005
  [arXiv:0809.3264 [hep-ph]].
  

\bibitem{Aad:2015iea}
  G.~Aad {\it et al.} [ATLAS Collab.],
  arXiv:1507.05525 [hep-ex].
  
\bibitem{Aad:2015zhl}
  G.~Aad {\it et al.} [ATLAS and CMS Collab.s],
  Phys.\ Rev.\ Lett.\  {\bf 114} (2015) 191803
  [arXiv:1503.07589 [hep-ex]].

\bibitem{Barbieri:1987fn}
J.R. Ellis, K. Enqvist, D. V. Nanopoulos and
F. Zwirner. Mod. Phys. Lett. {\bf A1} (1986) 57;
  R.~Barbieri and G.~F.~Giudice,
  Nucl.\ Phys.\ B {\bf 306} (1988) 63.

\bibitem{Feng:1999mn}
  J.~L.~Feng, K.~T.~Matchev and T.~Moroi,
  Phys.\ Rev.\ Lett.\  {\bf 84} (2000) 2322
  [hep-ph/9908309].
  
\bibitem{Kitano:2006gv}
  R.~Kitano and Y.~Nomura,
  Phys.\ Rev.\ D {\bf 73} (2006) 095004
  [hep-ph/0602096].

\bibitem{weiler_n}
M. Papucci, J.T. Ruderman and A. Weiler, JHEP {\bf 1209} (2012) 035
[arXiv:1110.6926 [hep-ph]].

\bibitem{Casas:2014eca}
  J.~A.~Casas, J.~M.~Moreno, S.~Robles, K.~Rolbiecki and B.~Zaldívar,
  JHEP {\bf 1506} (2015) 070
  [arXiv:1407.6966 [hep-ph]].

\bibitem{tev_tune}
H. Baer, V. Barger, P. Huang, A. Mustafayev and X. Tata,
Phys. Rev. Lett. {\bf 109} (2012) 161802 [arXiv:1207.3343 [hep-ph]],
and Phys. Rev. {\bf D87} (2013) 11, 115028 [arXiv:1212.2655 [hep-ph]].
    
\bibitem{Aad:2015pfx}
  G.~Aad {\it et al.} [ATLAS Collab.],
  arXiv:1506.08616 [hep-ex].
  
\bibitem{Chatrchyan:2013xna}
  S.~Chatrchyan {\it et al.} [CMS Collab.],
  Eur.\ Phys.\ J.\ C {\bf 73} (2013) 12,  2677
  [arXiv:1308.1586 [hep-ex]].
  

\bibitem{Baer:2012uy}
  H.~Baer, V.~Barger, P.~Huang and X.~Tata,
  JHEP {\bf 1205} (2012) 109 [arXiv:1203.5539 [hep-ph]].

\bibitem{Buchmueller:2013exa}
  O.~Buchmueller and J.~Marrouche,
  Int.\ J.\ Mod.\ Phys.\ A {\bf 29} (2014) 06,  1450032
  [arXiv:1304.2185 [hep-ph]].

\bibitem{Han:2013kga}
  C.~Han, K.-i.~Hikasa, L.~Wu, J.~M.~Yang and Y.~Zhang,
  JHEP {\bf 1310} (2013) 216
  [arXiv:1308.5307 [hep-ph]].

\bibitem{Belanger:2015vwa}
  G.~B\'elanger, D.~Ghosh, R.~Godbole and S.~Kulkarni,
  arXiv:1506.00665 [hep-ph].

\bibitem{Kobakhidze:2015scd}
  A.~Kobakhidze, N.~Liu, L.~Wu, J.~M.~Yang and M.~Zhang,
  arXiv:1511.02371 [hep-ph].
  
\bibitem{Kowalska:2013ica}
  K.~Kowalska and E.~M.~Sessolo,
  Phys.\ Rev.\ D {\bf 88} (2013) 7,  075001
  doi:10.1103/PhysRevD.88.075001
  [arXiv:1307.5790 [hep-ph]].
  
\bibitem{Gamberini:1989jw}
  G.~Gamberini, G.~Ridolfi and F.~Zwirner,
  Nucl.\ Phys.\ B {\bf 331} (1990) 331.

\bibitem{Gabbiani:1996hi}
  F.~Gabbiani, E.~Gabrielli, A.~Masiero and L.~Silvestrini,
  Nucl.\ Phys.\ B {\bf 477} (1996) 321
  [hep-ph/9604387].

\bibitem{loophole}
D.S.M. Alves, M.R. Buckley, P.J. Fox, J.D. Lykken and C.-T. Yu,
Phys. Rev. {\bf D87} (2013) 035016 [arXiv:1205.5805 [hep-ph]];
M.R. Buckley, T. Plehn and M.J. Ramsey-Musolf, Phys. Rev. {\bf D90}
(2014) 014046 [arXiv:1403.2726 [hep-ph]].

\bibitem{coann}
J.R. Ellis, T. Falk and K.A. Olive, Phys. Lett. {\bf B444} (1998) 367
[hep-ph/9810360].

\bibitem{baer_axi}
K.J. Bae, H. Baer and E.J. Chun, Phys. Rev. {\bf D89} (2014) 3, 031701
[arXiv:1309.0519 [hep-ph]].

\bibitem{gunion}
C.H. Chen, M. Drees and J.F. Gunion, Phys. Rev. {\bf D55} (1997) 330;
erratum, Phys. Rev. {\bf D60} (1999) 039901 [hep-ph/9607421].

\bibitem{higgsino_mono}
H. Baer, A. Mustafayev and X. Tata, Phys. Rev. {\bf D89} (2014) 055007
[arXiv:1401.1162 [hep-ph]].

\bibitem{Hikasa:1987db}
  K.-i.~Hikasa and M.~Kobayashi,
  Phys.\ Rev.\ D {\bf 36} (1987) 724.
 
\bibitem{Baer:1990sc}
  H.~Baer, X.~Tata and J.~Woodside,
  Phys.\ Rev.\ D {\bf 42} (1990) 1568.

\bibitem{Chalons:2015vja}
  G.~Chalons and D.~Sengupta,
  arXiv:1508.06735 [hep-ph].
  
\bibitem{Porod:2011nf}
  W.~Porod and F.~Staub,
  Comput.\ Phys.\ Commun.\  {\bf 183} (2012) 2458
  [arXiv:1104.1573 [hep-ph]].
  
\bibitem{Alwall:2011uj}
  J.~Alwall, M.~Herquet, F.~Maltoni, O.~Mattelaer and T.~Stelzer,
  JHEP {\bf 1106} (2011) 128
  [arXiv:1106.0522 [hep-ph]].
  
\bibitem{Sjostrand:2006za}
  T.~Sj\"ostrand, S.~Mrenna and P.~Z.~Skands,
  JHEP {\bf 0605} (2006) 026
  [hep-ph/0603175].

\bibitem{Sjostrand:2014zea}
  T.~Sj\"strand {\it et al.},
  Comput.\ Phys.\ Commun.\  {\bf 191} (2015) 159
  [arXiv:1410.3012 [hep-ph]].
  
\bibitem{Drees:2013wra}
  M.~Drees, H.~Dreiner, D.~Schmeier, J.~Tattersall and J.~S.~Kim,
  Comput.\ Phys.\ Commun.\  {\bf 187} (2014) 227
  [arXiv:1312.2591 [hep-ph]].

\bibitem{webpage}
https://checkmate.hepforge.org/

\bibitem{manager}
J.~S.~Kim, D.~Schmeier, J.~Tattersall and K.~Rolbiecki,
  Comput.\ Phys.\ Commun.\  {\bf 196} (2015) 535
  [arXiv:1503.01123 [hep-ph]].

\bibitem{deFavereau:2013fsa}
  J.~de Favereau {\it et al.}  [DELPHES 3 Collab.],
  JHEP {\bf 1402} (2014) 057
  [arXiv:1307.6346 [hep-ex]].

\bibitem{stable}
S. Chatrchyan {\it et al.} [CMS Collab.], JHEP {\bf 1307} (2013) 122
[arXiv:1305.0491 [hep-ex]];
G. Aad {\it et al.} [ATLAS Collab.], Eur. Phys. J. {\bf C75} (2015) 407
[arXiv:1506.05332 [hep-ex]];


\bibitem{Drees:1990dx}
  M.~Drees and K.~Hagiwara,
  Phys.\ Rev.\ D {\bf 42} (1990) 1709.
 
\bibitem{Abbiendi:2002vz}
  G.~Abbiendi {\it et al.}  [OPAL Collab.],
  Eur.\ Phys.\ J.\ C {\bf 29} (2003) 479
  [hep-ex/0210043].
  
\bibitem{Abdallah:2003xe}
  J.~Abdallah {\it et al.}  [DELPHI Collab.],
  Eur.\ Phys.\ J.\ C {\bf 31} (2003) 421
  [hep-ex/0311019].
  
\bibitem{Abbiendi:2003sc}
  G.~Abbiendi {\it et al.}  [OPAL Collab.],
  Eur.\ Phys.\ J.\ C {\bf 35} (2004) 1
  [hep-ex/0401026].
 
\bibitem{monojet}
M. Carena, A. Freitas and C. E. M. Wagner, JHEP {\bf 0810} (2008) 109
[arXiv:0808.2298 [hep-ph]]; 
M. Drees, M. Hanussek and J.S. Kim,  Phys. Rev. {\bf D86} (2012) 035024
[arXiv:1201.5714 [hep-ph]].

\bibitem{Mangano:2006rw}
  M.~L.~Mangano, M.~Moretti, F.~Piccinini and M.~Treccani,
  JHEP {\bf 0701} (2007) 013
  [hep-ph/0611129].

\bibitem{ATLAS-CONF-2012-097}
“Measuring the b-tag efficiency in a top-pair sample with 4.7 fb-1 of data from the ATLAS detector,” Tech. Rep. ATLAS-CONF-2012-097, CERN, Geneva, Jul 2012.

\bibitem{ATLAS-CONF-2012-043}
“Measurement of the b-tag Efficiency in a Sample of Jets Containing Muons with 5 fb1 of Data from the ATLAS Detector,”
Tech. Rep. ATLAS-CONF-2012-043, CERN, Geneva, Mar 2012.


\bibitem{Read:2002hq}
  A.~L.~Read,
  J.\ Phys.\ G {\bf 28} (2002) 2693.

\bibitem{Aad:2013ija}
  G.~Aad {\it et al.}  [ATLAS Collab.],
  JHEP {\bf 1310} (2013) 189
  [arXiv:1308.2631 [hep-ex]].
  
\bibitem{Aad:2014qaa}
  G.~Aad {\it et al.}  [ATLAS Collab.],
  JHEP {\bf 1406} (2014) 124
  [arXiv:1403.4853 [hep-ex]].
  
\bibitem{Lester:1999tx}
  C.~G.~Lester and D.~J.~Summers,
  Phys.\ Lett.\ B {\bf 463} (1999) 99
  [hep-ph/9906349].
  
\bibitem{Aad:2014pda}
  G.~Aad {\it et al.}  [ATLAS Collab.],
  JHEP {\bf 1406} (2014) 035
  [arXiv:1404.2500 [hep-ex]].
  
\bibitem{Aad:2014kra}
  G.~Aad {\it et al.}  [ATLAS Collab.],
  JHEP {\bf 1411} (2014) 118
  [arXiv:1407.0583 [hep-ex]].
  
\bibitem{Aad:2014nra}
  G.~Aad {\it et al.}  [ATLAS Collab.],
  Phys.\ Rev.\ D {\bf 90} (2014) 5,  052008
  [arXiv:1407.0608 [hep-ex]].
  
\bibitem{Chatrchyan:2013lya}
  S.~Chatrchyan {\it et al.}  [CMS Collab.],
  Eur.\ Phys.\ J.\ C {\bf 73} (2013) 9,  2568
  [arXiv:1303.2985 [hep-ex]].
  
  \bibitem{ATLAS-CONF-2012-104}
G. Aad {\it et al.} [ATLAS Collab.], Tech. Rep. ATLAS-CONF-2012-104,
CERN, Geneva, Aug 2012. 

\bibitem{ATLAS-CONF-2013-024}
G. Aad {\it et al.} [ATLAS Collab.], Tech. Rep. ATLAS-CONF-2013-024,
CERN, Geneva, Mar 2013. 

\bibitem{ATLAS-CONF-2013-047}
G. Aad {\it et al.} [ATLAS Collab.], Tech. Rep. ATLAS-CONF-2013-047,
CERN, Geneva, May 2013.

\bibitem{ATLAS-CONF-2013-061}
G. Aad {\it et al.} [ATLAS Collab.], Tech. Rep. ATLAS-CONF-2013-061,
CERN, Geneva, Jun 2013.

\bibitem{ATLAS-CONF-2013-062}
G. Aad {\it et al.} [ATLAS Collab.], Tech. Rep. ATLAS-CONF-2013-062,
CERN, Geneva, Jun 2013.

\bibitem{CMS-PAS-SUS13-016}
S. Chatrchyan {\it et al.} [CMS Collab.], CMS-PAS-SUS13-016, CERN,
Geneva, November, 2013.

\bibitem{s95}
See e.g. the chapter on Statistics in K.A. Olive et al. (Particle Data
Group), Chin. Phys. {\bf C38} (2014) 090001.



\end{thebibliography}

 
\end{document}